\documentclass[prl,twocolumn,superscriptaddress,citeautoscript,amsart,longbibliography]{revtex4}

\usepackage{subfiles}
\usepackage{appendix}
\usepackage{graphicx, tikz-cd}
\usepackage{multirow}
\usepackage{xcolor}
\usepackage{bm}
\usepackage{times}
\usepackage{amsmath,bm,amsfonts}
\usepackage{dcolumn}
\usepackage{graphicx}
\usepackage{latexsym}
\usepackage{bbm,amsthm,amssymb,verbatim,tikz,physics}

\usepackage{ulem} 
\usepackage{braket}
\usepackage{bbold}

\def\ket#1{|#1\rangle }
\def\bra#1{\langle #1 |}
\def\n{\nonumber \\ }

\def\bb{\mathbb}

\def\Re{\mathrm{Re}}

\def\ket#1{|#1\rangle }
\def\bra#1{\langle #1 |}
\def\n{\nonumber \\ }

\def\bb{\mathbb}
\def\bf{\mathbf}

\def\Re{\mathrm{Re}}
\def\eq#1{Eq.~\eqref{#1}}
\def\bk{\mathbf{k}}

\begin{document}
\title{Unconventional topological phase transition of the Hopf insulator}

\author{Sunje \surname{Kim}}\thanks{These authors contributed equally to this work.}
\affiliation{Department of Physics and Astronomy, Seoul National University, Seoul 08826, Korea}
\affiliation{Center for Theoretical Physics (CTP), Seoul National University, Seoul 08826, Korea}
\affiliation{Institute of Applied Physics, Seoul National University, Seoul 08826, Korea}

\author{Ysun \surname{Choi}}\thanks{These authors contributed equally to this work.}
\affiliation{Department of Physics and Astronomy, Seoul National University, Seoul 08826, Korea}
\affiliation{Center for Theoretical Physics (CTP), Seoul National University, Seoul 08826, Korea}
\affiliation{Institute of Applied Physics, Seoul National University, Seoul 08826, Korea}

\author{Hyeongmuk \surname{Lim}}\thanks{These authors contributed equally to this work.}
\affiliation{Department of Physics and Astronomy, Seoul National University, Seoul 08826, Korea}
\affiliation{Center for Theoretical Physics (CTP), Seoul National University, Seoul 08826, Korea}
\affiliation{Institute of Applied Physics, Seoul National University, Seoul 08826, Korea}

\author{Bohm-Jung \surname{Yang}}
\email{bjyang@snu.ac.kr}
\affiliation{Department of Physics and Astronomy, Seoul National University, Seoul 08826, Korea}
\affiliation{Center for Theoretical Physics (CTP), Seoul National University, Seoul 08826, Korea}
\affiliation{Institute of Applied Physics, Seoul National University, Seoul 08826, Korea}

\date{\today}

\begin{abstract}
The topological phase transition between two band insulators is mediated by a gapless state whose low energy band structure normally contains sufficient information for describing the topology change.
In this work, we show that there is a class of topological insulators whose topological phase transition cannot be explained by this conventional paradigm. 
Taking the Hopf insulator as a representative example, we show that the change of the Hopf invariant requires the information of wave functions as well as the gapless band structure simultaneously. 
More explicitly, the description of the Hopf invariant change requires us to trace not only the trajectory of Weyl points but also the evolution of the preimages for two distinct eigenstates.
We show that such an unconventional topological phase transition originates from the fact that the Hopf invariant is well-defined when all the lower dimensional topological invariants are trivial, which in turn allows us to lift the classifying space of occupied state projectors to the corresponding universal 2-covering space of wave functions. 
Generalizing our theory to inversion-symmetric 10-fold Altland-Zirnbauer symmetry classes, we provide a complete list of symmetry classes, all of which turn out to have delicate band topology related to the Hopf invariant, where similar unconventional topological phase transitions can appear.
\end{abstract}

\maketitle
\def\thefootnote{*}\footnotetext{These authors contributed equally to this work}

\textit{Introduction.|}
The topological phase transition (TPT) of an insulator is accompanied by an accidental band crossing (ABC) which generates various gapless nodes~\cite{reviewTI,QSHI,bradlyn2017topological,
ClassificationDSM,bouhon2020nonabelian,
ahn2018band,bzdusek2020nonabelian,
ahn2019stiefel,FermiarcVishwanath}.
In three-dimensional (3D) systems without any local symmetry, belonging to the class A of the Altland-Zirnbauer (AZ) classification,
an ABC between nondegenerate bands generally creates Weyl points (WPs)~\cite{FermiarcVishwanath,xu2015observation, FermiarcExperiment,burkov2012chiralanomaly,reviewWSM}. The momentum space trajectory traversed by WPs during the transition between two insulators forms a closed loop whose geometric shape characterizes the nature of the TPT
\cite{reviewWSM,ahn2017unconventional,
murakami2008Z2TIphasetransition}.  For instance, in the transition from a normal insulator (NI) to a 3D Chern insulator (CI), the trajectory of WPs forms a non-contractible loop along a momentum direction. 
This is because, since a WP carries a Chern number (CN), two WPs with opposite CNs should traverse the entire Brillouin zone (BZ) along a momentum direction to transform a NI into a 3D CI. This also indicates that if the WP trajectory forms a contractible loop between two insulators, their CNs should be identical. 

The Hopf insulator (HI) is another type of 3D topological insulators appearing in time-reversal broken two-band systems with various intriguing properties~\cite{nelson2021multicellularity,alex2021hopfsurface,
penghao2021quantized,nelson2022delicate,
moore2021realizinghopf,deng2013hopf} 
The corresponding topological invariant, the Hopf invariant, is a genuine 3D topological number distinct from the CN.
Since the Hopf invariant is gauge-invariant when the CN vanishes, the HI is well-defined under zero CN condition~\cite{wen2008hopf}. 
Since an ABC is generally accompanied by WP creations in the HI as well, the interplay of the Hopf invariant and WP trajectory is critical in describing the Hopf invariant changing TPT. However, its general understanding is still missing.

\begin{figure}[t]
	\includegraphics[width=\linewidth]{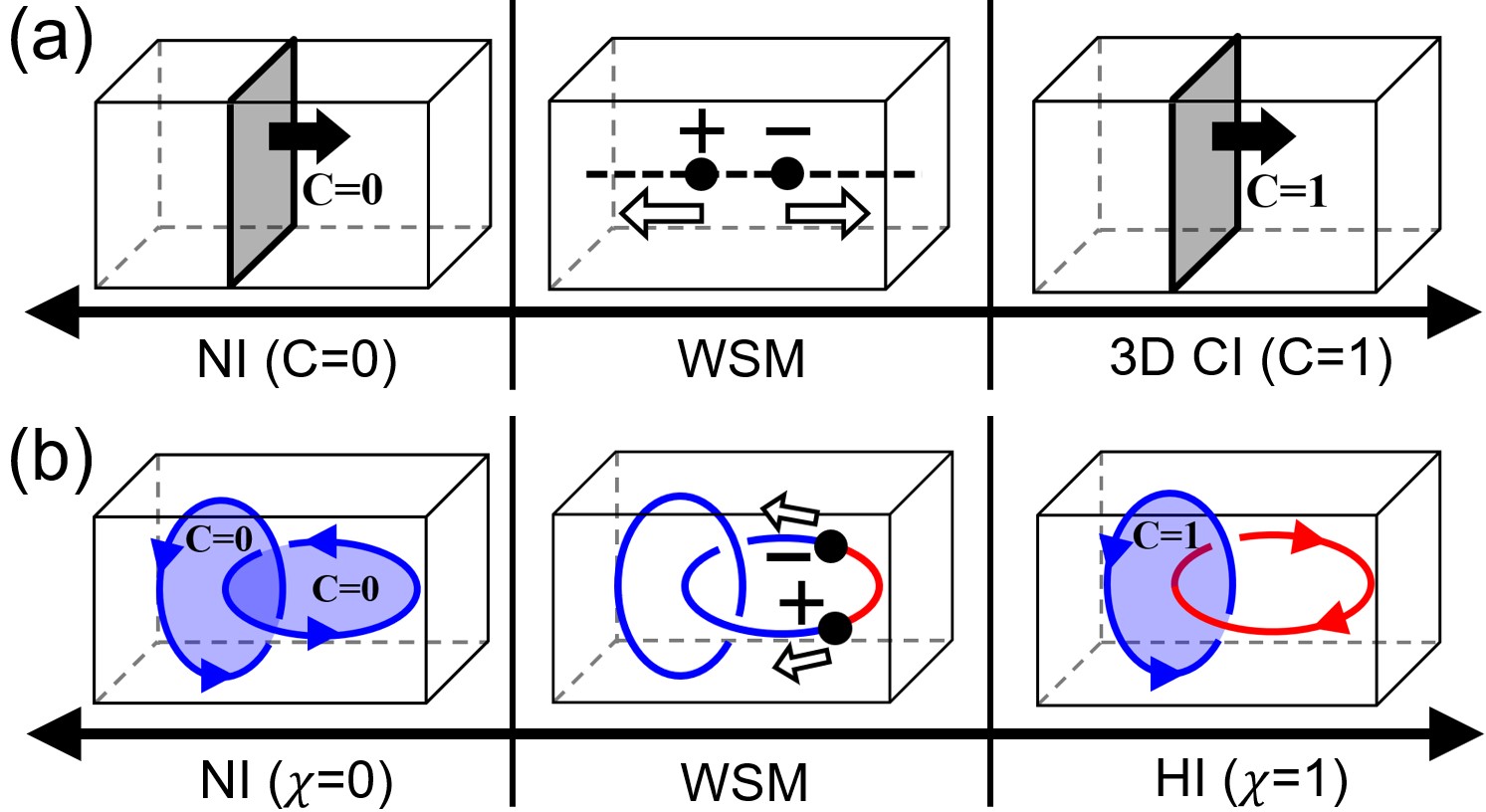}
	\caption{
		Schematic topological phase diagram for 3D magnetic systems without local symmetry.
		(a) Transiton from a normal insulator (NI) to a 3D Chern insulator (CI) mediated by Weyl semimetal (WSM).
		Weyl point (WP) trajectory forms a noncontractible loop across the Brillouin zone (BZ).
		(b) Transiton from a NI to a Hopf insulator (HI) in which the WP trajectory forms a contractible loop.
		WPs are depicted by black dots, which change the Chern number (CN) by passing through the blue surface. The red and blue lines are preimages for different eigenstates.
	}
	\label{fig:CIHI}
\end{figure}

In this letter, we construct a general theory for the TPT between two insulators with distinct Hopf invariants mediated by a Weyl semimetal (WSM) phase in between.
If the two insulators satisfy the zero CN condition, the WP trajectory in the intermediate WSM should form a contractible loop. Thus, the Hopf invariant changing TPT is distinct from the CN changing TPT as compared in Fig.~\ref{fig:CIHI}. 
Interestingly, contrary to any known TPT studied before, the Hopf invariant changing TPT requires the information of the wave function and the WP trajectory simultaneously.
More specifically, the HI Hamiltonian describes a map from $\mathbb{T}^{3}_{\text{BZ}}$ to $\mathbb{S}^{2}$ where the former (latter) indicates the 3D BZ torus (Bloch sphere (BS)).
For a given point on the BS representing an occupied state, the corresponding preimages form 1D loops in the 3D BZ. 
To describe the Hopf invariant changing TPT, we need to examine the evolutions of the preimages relevant to two distinct points on the BS as well as the WP trajectory.
We show that the fundamental origin of such an unconventional TPT originates from the fact that the Hopf invariant is well-defined when the CNs are trivial, which in turn allows us to lift the classifying space of occupied state projectors to the corresponding universal 2-covering space of wave functions.  
Since the wave function space have more information than the classifying space, the HI TPT occurs under stricter conditions leading to the unconventional TPT.
We generalize the discussion to inversion-symmetric 10-fold AZ symmetry classes \cite{bzdusek2017robust} and show that the unconventional TPT is deeply related to the delicate topology of Hopf invariant.

\textit{Hopf insulator and lifting.|}
A general two-band Hamiltonian without any local symmetry can be written as
\begin{gather}
    H_g(\bk) = f_1(\bk)\sigma_{x} + f_2(\bk)\sigma_{y} + f_3(\bk)\sigma_{z}\equiv \mathbf{f}(\bk)\cdot\mathbf{\sigma},
\label{eq.2band}
\end{gather}
where $f_{1,2,3}(\bk)$ are real functions of the momentum $\bk$ and $\sigma_{x,y,z}$ are Pauli matrices.
The term related to the chemical potential is not included as it does not affect the band topology.
The gapped band topology of $H_g(\bk)$ can be fully described by the normalized Bloch vector $\hat{\mathbf{f}}(\bk)\equiv\mathbf{f}(\bk)/|\mathbf{f}(\bk)|\in\mathbb{S}^{2}$ on the BS $\mathbb{S}^{2}$, equivalent to the classifying space $ \mathfrak{X}= \frac{\mathrm{U}(2)}{\mathrm{U}(1)\times \mathrm{U}(1)} \simeq \mathbb{S}^{2}$ of the occupied state projector $P(\bk)=\ket{\psi(\bk)}\bra{\psi(\bk)}$ where $\ket{\psi(\bk)}$ indicates the occupied state.
Thus, any 3D two-band insulator can be identified with a continuous map $\hat{f}: \mathbb{T}^{3}_{\text{BZ}} \to \mathbb{S}^{2}$.
Three gap-closing equations $f_{1,2,3}(\mathbf{k},\phi)=0$ give 1D gapless solutions in the four-dimensional (4D) space $(\mathbf{k},\phi)$, which indicates an insulator-WSM transition when an ABC occurs by changing a control parameter $\phi$~\cite{murakami2008Z2TIphasetransition}.

Under the zero CN condition, the HI Hamiltonian can be written as \cite{wen2008hopf}
\begin{subequations}
	\begin{gather}
		H_{\text{Hopf}}(\bk) = -[\tilde{f}^{\dagger}(\bk) \bm{\sigma} \tilde{f}(\bk)] \cdot\bm{\sigma}, 
		\label{eq.hopf}
	\end{gather}
\end{subequations}
where the complex two-component vector $\tilde{f}(\bk)\in\mathbb{C}^{2}$ indicates the occupied eigenstate of $H_{\text{Hopf}}(\bk)$.
Imposing the normalization condition $\|\tilde{f}\|^2 = 1$, we have $\tilde{f}(\bk)\in \mathbb{S}^{3}$.
Thus, $H_{\text{Hopf}}(\bk)$ corresponds to a two-step mapping $\mathbb{T}^{3}_{\text{BZ}} \to \mathbb{S}^3 \rightarrow \mathbb{S}^2$ that ``factors through" the 3-sphere $\mathbb{S}^{3}$, while $H_g(\bk)$ corresponds to a single-step mapping $\mathbb{T}^3_{\text{BZ}} \rightarrow \mathbb{S}^2$. 
The reason behind such difference is that the wave function can be defined continuously over the entire 3D BZ under zero CN condition that gives the deformation map $\tilde{f}:\mathbb{T}^{3}_{\text{BZ}} \to \mathbb{S}^{3}$.
The second map corresponds to the Hopf fibration $p: \mathbb{S}^{3} \to \mathbb{S}^{2}$ described by
\begin{gather}
	p: \tilde{f} \mapsto -\tilde{f}^{\dagger} \bm{\sigma} \tilde{f} = (f_x, f_y, f_z),
\end{gather}
which discards the phase information as $p(\tilde{f}e^{i\theta}) = p(\tilde{f})$.
Thus, it can be interpreted as the projector function
\begin{gather}
	p: \ket{\psi}\in \mathbb{S}^{3} \mapsto \ket{\psi}\bra{\psi} = P(\bk)\in \mathbb{S}^{2}.
\end{gather}

The situation is succinctly summarized by the notion of \textit{lifting}:
\begin{gather}
	\label{cd.hopf}
	\begin{tikzcd}[ampersand replacement=\&] 
		\& \quad\tilde{\mathfrak{X}}=\mathbb{S}^{3} \dar["p"] \\
		\mathbb{T}^{3}_{\text{BZ}} \ar[ur,"\tilde{f}"] \rar["f"] \& \quad\mathfrak{X}=\mathbb{S}^{2}
	\end{tikzcd}
\end{gather} 
While $H_g(\bk)$ is represented by a map $f: \mathbb{T}^{3}_{\text{BZ}} \to \mathbb{S}^{2}$, the HI enjoys the lifting of the map $f$ to another map $\tilde{f}$. That $\tilde{f}$ is a lift of $f$ is equivalent to saying that $p\circ\tilde{f} = f$.
Since the Hopf fibration $p$ maps a circle $\mathbb{S}^{1}_{x}$ onto a single point $x\in\mathbb{S}^{2}$, finding a lift $\tilde{f}$ of a general map $f$ is equivalent to choosing a point $\widetilde{f}(\bk) \in \mathbb{S}^{1}_{f(\bk)}$ coherently for all momenta $\bk\in\mathbb{T}^{3}_{\text{BZ}}$. The obstruction for finding such a lift is precisely the nonzero CNs on any 2D subtori of $\mathbb{T}^{3}_{\mathrm{BZ}}$.

We note that the gap-closing condition for $H_{\text{Hopf}}$ is given by the following four independent equations
\begin{gather}
	\Re[\tilde{f}(\bk,\phi)] = 0, \quad
	\Im[\tilde{f}(\bk,\phi)] = 0, 
	\label{eq.4eqs}
\end{gather}
which can be contrasted to the three equations $f_{1,2,3}(\mathbf{k},\phi)=0$ of $H_g(\bk)$.
Hence, the TPT of $H_{\text{Hopf}}$ is expected to occur through a single critical point in the 4D space  $(\mathbf{k},\phi)$,
which implies that the WP trajectory is not sufficient to describe the Hopf invariant change.
In fact, when small perturbation is added to $H_{\text{Hopf}}$, an ABC accompanies an insulator-WSM transition. 
However, unlike general TPT in class A, describing the Hopf invariant change involves more information (individual states $\ket{\psi(\bk)}$) than WP trajectory (encoded in $\ket{\psi(\bk)}\bra{\psi(\bk)}$).

\textit{Hopf invariant.|}
The Hopf invariant $\chi$ can be computed by using the preimages $\gamma_y=\hat{\mathbf{f}}^{-1}[{y}]$ of a point $y\in\mathbb{S}^2$ on the BS. 
Generically, $\gamma_y$ is a collection of loops $\gamma_{y,i}$ with the index $i$ in the 3D BZ . 
As the Berry curvature vector at each point of $\gamma_{y,i}$ is always parallel to $\gamma_{y,i}$ \cite{SMHIPT},
the orientation of $\gamma_{y,i}$ can be well-defined as the opposite to the Berry curvature direction. 
Using the oriented loops $\{\gamma_{i}\}$ with the relevant oriented surfaces $\{\Sigma_{i}\}$ 
bounded by $\{\gamma_{i}\}$, $\chi$ can be written by the Whitehead's integral formula~\cite{whitehead1947expression,
nelson2021multicellularity}
\begin{gather}
    \chi=\sum_{i}{{1 \over 2\pi}\int_{\Sigma_{i}}  \bf{F}\cdot d{\boldsymbol\Sigma}},
\label{eq.Whitehead}
\end{gather}
which can be interpreted as the sum of the CN $C_{i}$ on each surface $\Sigma_{i}$ whose boundary $\gamma_{i}$ is topologically contractible to a point on the BS.
We note that the Whitehead's formula allows us to describe $\chi$ as the linking number between two sets of preimages $\gamma=\cup_{i}\gamma_{i}$ and $\gamma'=\cup_{i}\gamma'_{i}$ for two different points $y$ and $y'$ on the BS.
This is because $\gamma'$ should pierce $\Sigma_{i}$ $C_{i}$ times (as much as its CN).  
Since $\gamma'$ is always antiparallel to the Berry curvature direction, 
counting the occurrence of $\Sigma_{i}$ pierced by $\gamma'$ considering the orientation of $\gamma_{i}$ is equivalent to counting the linking number between $\gamma$ and $\gamma'$ \cite{SMHIPT}.

\begin{figure}[t]
	\includegraphics[width=\linewidth]{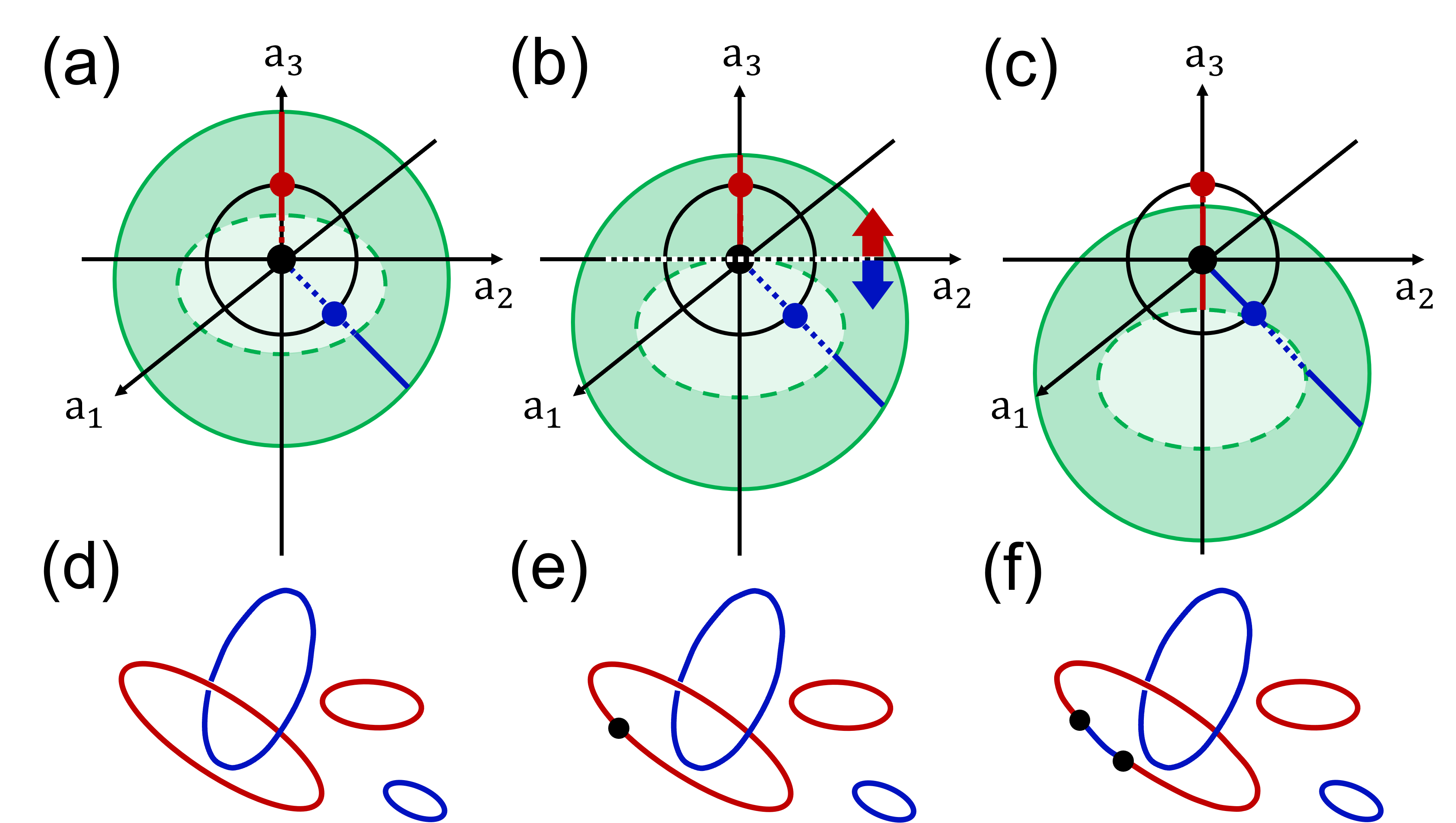}
	\caption{Schematics of (a,b,c) the Bloch space $\mathbf{f}[\mathbb{T}^{3}_{\text{BZ}}]$ and (d,e,f) the preimages as the HI transforms to the WSM. 
	(a) $\mathbf{f}[\mathbb{T}^{3}_{\text{BZ}}]$ of a HI with an inner cavity (dashed circle) enclosing the origin. 
	The blue and red solid lines indicate the 1D segments in the Bloch space that project onto the blue and red dots on the BS (black circle), respectively, whose preimages are depicted in (d) with the same colors. 
	(b) Similar figure at the critical point of the insulator-semimetal transition.
	The surface of the cavity touches the origin.
	The gapless point (black dot in (e)), which is the inverse image of the origin, is generated and attached to the red preimage, as depicted in (e). 
	(c) In the WSM, the origin is inside the bulk of $\mathbf{f}[\mathbb{T}^{3}_{\text{BZ}}]$. 
	Both the red and blue solid lines are attached to the origin, indicating that their preimages are also attached to the WPs (black dots), as illustrated in (f).
	}
	\label{Fig3_0420.png}
\end{figure}

\textit{Preimages and Weyl points.|}
Let us examine the relation between the preimages of $\hat{\mathbf{f}}(\bk)$ and WPs in the WSM.
Since a generic WP carries a unit CN, the image of a small $\epsilon$-sphere centered at the WP should wind the BS exactly once, giving one-to-one correspondence between the $\epsilon$-sphere and the BS. 
Even in the WSM, by excluding small $\epsilon$-spheres of all WPs, one can track the preimages of the BS. Generally, the preimage $\gamma$ in the WSM is composed of closed loops as in insulators plus open lines. Here, a tail of a open line must be attached to a WP due to the correspondence between the $\epsilon$-sphere and the BS.

For more systematic description, let us introduce the {\it Bloch space} spanned by the unnormalized Bloch vector $\mathbf{f}=(f_1,f_2,f_3)$. 
The origin $\mathbf{f}=(0,0,0)$ represents the gap-closing of the Hamiltonian. 
The image $\mathbf{f}[\mathbb{T}^3_{\text{BZ}}]$ of an insulator is a 3D object with an inner cavity enclosing the origin whose projection onto the unit sphere gives the BS [Fig.~\ref{Fig3_0420.png}(a,d)]. 
According to the Whitehead's formula in Eq. (\ref{eq.Whitehead}), 
when $\chi\neq0$, $\mathbf{f}[\mathbb{T}^3_{\text{BZ}}]$ should wrap around the origin because at least one $\Sigma_{i}$ has a non-trivial CN.

As an insulator gets close to a WSM, generally there is a particular direction that the surface of the cavity in $\mathbf{f}[\mathbb{T}^{3}_{\text{BZ}}]$ approaches the origin. 
At the critical point, the BS can be demarcated by two hemispheres with the border plane perpendicular to the approaching direction. 
At this moment, the gap-closing points which are the preimage of the origin should be generated on the preimages $\gamma_{y}$ for $y$ located in the hemisphere without the cavity [the upper hemisphere in Fig. \ref{Fig3_0420.png} (b)]. 
Namely, all the preimages $\gamma_{y}$ for the points $y$ located on this hemisphere must be attached to the gap-closing point [Fig. \ref{Fig3_0420.png} (e)]. 
Whereas, the gap-closing point does not touch the preimages $\gamma_{y'}$ for $y'$ located on the other hemisphere [the lower hemisphere in Fig. \ref{Fig3_0420.png} (b)]. 

When the system evolves into the WSM and the origin gets immersed in the bulk of $\mathbf{f}\left[\mathbb{T}^{3}_{\text{BZ}}\right]$, every preimage has its image in the vicinity around the origin, and thus should be connected to WPs. 
Namely, the preimage $\gamma_{y,i}$, which was a closed loop attached to the gap-closing point at the critical point, becomes an open line with two WPs at its boundary [Fig. \ref{Fig3_0420.png} (f)]. 
Whereas, for $y'$ located on the other hemisphere, a preimage segment attached to WPs should be newly generated in the WSM [Fig. \ref{Fig3_0420.png} (f)].
A similar idea can also be applied to the WSM-to-insulator transition.

\textit{TPT changing the Hopf invariant.|}
To describe the Hopf invariant changing TPT, one needs to trace not only the WP trajectory, but also the preimages of two different points on the BS.
Even when the WP trajectory and the preimage evolution of one point on the BS are the same, the Hopf invariant change $\Delta\chi$ can be different \cite{SMHIPT}.

\begin{figure}[t]
\includegraphics[width=\linewidth]{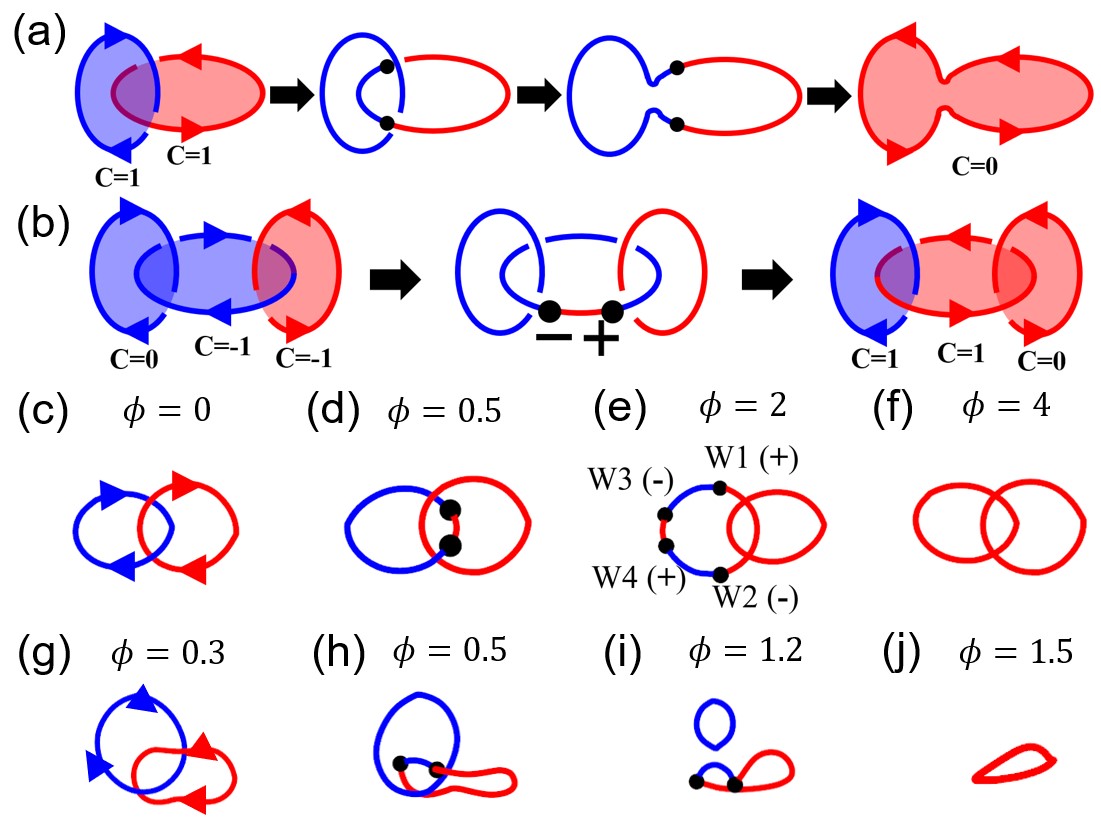}
\caption{
Schematic figures for loop annihilation processes and TPTs of lattice models.
(a) Merging process between a loop and an open line. 
(b) Destruction process by pair-created WPs consuming the blue loop. The blue and red lines are the preimages of two different points on the BS and black dots are WPs.
(c)-(f) WPs and preimages of the points on the BS for the model in Eq.~(\ref{WPT1}) with $\phi\in[0,4]$ and $\theta=1/2$.
The blue (red) loop is the preimage of $(-1,0,0)$ ($(1,0,0)$) on the BS.
The directions of the blue and red loops are given by the arrows on the lines.
In (e), WPs are labelled from W1 to W4 and their charges are written in the bracket.
(g)-(j) TPT for the model in Eq.~(\ref{WPT1}) with $\theta=\phi\in[0,3/2]$.
The red (blue) loop is the preimage of $(0,\frac{\sqrt{3}}{2},-\frac{1}{2})$ ($(1,0,0)$) on the BS.
}
\label{fig:merging_consuming}
\end{figure}

To compute $\Delta\chi$, we classify the preimage loops of a point on the BS into three types: generated, intact, and disappearing loops across the WSM phase.
Then, $\Delta\chi$ becomes
\begin{gather}
	\Delta\chi =\Delta C_{i} -C_{d}+C_{g},
	\label{Delchi}
\end{gather}
where $\Delta C_i$ is the CN change of intact loops and $C_d$ ($C_g$) is the CN of disappering (generated) loops.
We note that there are two different ways to obliterate a loop. 
One is through merging with another loop or an open line.
Since merging with another loop does not change the Hopf invariant, we only consider the latter [Fig.~\ref{fig:merging_consuming}(a)].
We note that merging can happen only between the preimages from the same point on the BS \cite{SMHIPT}.
The second way is through creating and annihilating WP pairs on a loop as illustrated in Fig. \ref{fig:merging_consuming} (b).

To describe $\Delta\chi$, we choose two points (point 1 and 2) on the BS and their preimages marked by blue and red colors, respectively.
First, to compute $\Delta C_{i}$, we consider the blue preimages of the point 1.
Then, $\Delta C_{i}=\Delta C_{1,i}$, the CN change on the surface bounded by the blue intact loops, induced by WP crossings.


Next, we consider $C_d$ of disappearing loops, either by merging with an open line or by being consumed by pair-created WPs.
First, when a loop merges with a open line as in Fig. \ref{fig:merging_consuming} (a), its linking number disappears, and thus we obtain
$\Delta\chi =\Delta L_{\mathrm{merge}}$, the linking number change between the merged loops before and after merging.
Second, for the blue-to-red transformation of a loop by pair-created WPs as in Fig. \ref{fig:merging_consuming} (b),
the linking number of the disappearing blue (generated red) loop will be subtracted (added).
Here, the points $1$ and $2$ are on the opposite side of the BS so that WPs are pair-created only on the preimages of the point 1.
The linking number of the generated red loop was already counted by $\Delta C_{i,1}$ because it is equal to the linking number change of the intact blue loops.
Similarly, the linking number of the disappearing blue loop can be counted by the linking number change of the intact red loops given by $\Delta C_{2,i}$, the CN change of the intact red preimage loop of the point $2$. Since a way to generate a loop can always be obtained by reversing the process to obliterate a loop, $C_{g}$ can be calculated exactly the same as $C_{d}$.

To sum up, $\Delta\chi$ can be expressed by 
\begin{gather}
\Delta\chi =\Delta C_{1,i}+\Delta C_{2,i} +\Delta L_{\mathrm{merge}},
\label{Delchi2}
\end{gather}
where $\Delta C_{1,i}$ are $\Delta C_{2,i}$ are the CN changes of intact loops and $\Delta L_{\mathrm{merge}}$ is the linking number change due to the merged or splitted loops during TPT.
This equation clearly shows that the TPT description requires not only tracking the WP trajectory but also tracing the preimage evolution of two different points on the BS.

To illustrate the TPT described by Eq. (\ref{Delchi2}), let us consider a modified Moore-Ran-Wen (MRW) model \cite{wen2008hopf}
\begin{gather}
	H_{\text{MRW}}=\mathbf{f}\cdot\boldsymbol{\sigma},~~\mathbf{f}=z^{\dagger}\boldsymbol{\sigma}z+\phi\hat{x},
	\\
	z=\begin{pmatrix}
		\sin k_{x} +i\sin k_{y} 
		\\ \sin{k_{z}}+i\left(\sum_{i=x,y,z} \cos{k_{i}} +2+\theta\right)
	\end{pmatrix},
	\label{WPT1}
\end{gather}
where $\phi$ and $\theta$ are constants. 
First, we change $\phi\in[0,4]$ for fixed $\theta=1/2$.
At $\phi=0$, the blue and red loops are linked as illustrated in Fig. \ref{fig:merging_consuming} (c).
Considering their directions illustrated by the arrows, we find that the initial Hopf invariant is $1$.
Around $\phi=0.5$ and $\phi=2$, pairs of WPs are created on the blue loop.
Increasing $\phi$ shifts the WP positions whose evolution transforms the blue lines into the red lines. Eventually, at $\phi=4$ where all WPs are pair-annihilated, a red loop is created completely.
Since the linking between the red and blue loops disappears through this process, we obtain $\Delta\chi=-1$.
In terms of Eq. (\ref{Delchi2}), since there is no merging or splitting of loops, we only have to consider the intact red loop. One can easily confirm that $\Delta C_{2,i}=\Delta\chi=-1$ by comparing the orientation of the surface bounded by the red loop and the WP charge crossing the surface.

As a second example, we change $\theta=\phi\in[0,3/2]$ as illustrated in Fig. \ref{fig:merging_consuming} (g-j).
Since there is no intact loop, $\Delta\chi$ is given by the linking number change of the split blue loop that is equal to the initial linking number at $\phi=0.3$, which gives $\Delta\chi=-1$.

{\it Discussion.|}
Unlike the conventional TPTs described by the trajectories of gapless nodes, it is crucial to track the eigenstate evolution for the HI TPT.
As illustrated in Eq.~(\ref{cd.hopf}), its fundamental origin is deeply related to the fact that $f: \mathbb{T}^{3}_{\text{BZ}} \to \mathfrak{X}=\mathbb{S}^{2}$ can be lifted to  $\tilde{f}: \mathbb{T}^{3}_{\text{BZ}} \to \tilde{\mathfrak{X}}=\mathbb{S}^{3}$ under zero CN condition, which also indicates the trivial induced homomorphism $f_{*}$ of the first and second homotopy groups, i.e.,
\begin{gather}
	0 = f_{*}: \pi_i[\mathbb{T}^{3}_{\text{BZ}}] \rightarrow \pi_i[\mathfrak{X}] \quad \text{for all } i\leq 2.
\end{gather}
The lifted classifying space $\tilde{\mathfrak{X}}=\mathbb{S}^{3}$ is called a universal 2-covering space of the classifying space $\mathfrak{X}=\mathbb{S}^{2}$.
In general, the map $f$ admits a lift $\tilde{f}$ that maps the BZ to a universal $n$-covering space $\tilde{\mathfrak{X}}_n$ of $\mathfrak{X}$ if and only if $f$ induces trivial homomorphism on all homotopy groups $\pi_i$ for $i\leq n$ \cite{MR1867354}.

Since a universal 1-covering space $\tilde{\mathfrak{X}}_1$ of $\mathfrak{X}$ generally has the same dimension as $\mathfrak{X}$, the presence of $\tilde{\mathfrak{X}}_2$ is crucial to observe the unconventional TPT that is achievable only in 3D systems.
For the HI, $\tilde{\mathfrak{X}}_2$ is actually the space of Bloch eigenstates, while $\mathfrak{X}$ itself is the space of occupied state projectors. Thus, $\tilde{\mathfrak{X}}_{2}$ has larger dimension than $\mathfrak{X}$, carrying more information.

We generalize the theory to all inversion-symmetric AZ symmetry classes whose classifying spaces generally take the form of
$\mathfrak{X} = G/H$ where $G$ is the configuration space of Bloch eigenstates and $H$ is the gauge group that describes mixing among the occupied (unoccupied) eigenstates~\cite{bzdusek2017robust}.
Using the long exact sequence of homotopy groups, we classify all possible cases that $G$ is a universal 2-covering space of $\mathfrak{X}=G/H$, which gives the complete list of 3D topological insulators with unconventional TPT when $G$ is the eigenstate space as follows \cite{wen2008hopf,SMHIPT, lim2023realhopf}:
i) class A (m=n=1), ii) class AI (\{m,n\}=\{1,2\},\{2,2\}), iii) class C (m=n=1), iv) class CI (m=n=2) 
where $m$ ($n$) is the number of unoccupied (occupied) bands.
Interestingly, the topological invariants of all these classes are the Hopf invariant \cite{SMHIPT}.
Also, as there is a strong constraint on the number of both occupied and unoccupied bands, we conclude that the unconventional TPT originates from the delicate topology of Hopf invariant.

\begin{acknowledgments}
S.K., Y.C., H.L., and B.-J.Y. were supported by Samsung Science and Technology Foundation under Project No. SSTF-BA2002-06, National Research Foundation of Korea (NRF) grants funded by the government of Korea (MSIT) (Grants No. NRF-2021R1A5A1032996), and GRDC(Global Research Development Center) Cooperative Hub Program through the National Research Foundation of Korea(NRF) funded by the Ministry of Science and ICT(MSIT) (RS-2023-00258359).
\end{acknowledgments}

\end{document}


\title{Supplementary Material for ``Unconventional topological phase transition of the Hopf insulator"}

\author{Sunje \surname{Kim}}\thanks{These authors contributed equally to this work.}
\affiliation{Department of Physics and Astronomy, Seoul National University, Seoul 08826, Korea}
\affiliation{Center for Theoretical Physics (CTP), Seoul National University, Seoul 08826, Korea}
\affiliation{Institute of Applied Physics, Seoul National University, Seoul 08826, Korea}

\author{Ysun \surname{Choi}}\thanks{These authors contributed equally to this work.}
\affiliation{Department of Physics and Astronomy, Seoul National University, Seoul 08826, Korea}
\affiliation{Center for Theoretical Physics (CTP), Seoul National University, Seoul 08826, Korea}
\affiliation{Institute of Applied Physics, Seoul National University, Seoul 08826, Korea}

\author{Hyeongmuk \surname{Lim}}\thanks{These authors contributed equally to this work.}
\affiliation{Department of Physics and Astronomy, Seoul National University, Seoul 08826, Korea}
\affiliation{Center for Theoretical Physics (CTP), Seoul National University, Seoul 08826, Korea}
\affiliation{Institute of Applied Physics, Seoul National University, Seoul 08826, Korea}

\author{Bohm-Jung \surname{Yang}}
\email{bjyang@snu.ac.kr}
\affiliation{Department of Physics and Astronomy, Seoul National University, Seoul 08826, Korea}
\affiliation{Center for Theoretical Physics (CTP), Seoul National University, Seoul 08826, Korea}
\affiliation{Institute of Applied Physics, Seoul National University, Seoul 08826, Korea}

\date{\today}

\maketitle
\def\thefootnote{*}\footnotetext{These authors contributed equally to this work}

\setcounter{section}{0}
\setcounter{figure}{0}
\setcounter{equation}{0}

\renewcommand{\thefigure}{S\arabic{figure}}
\renewcommand{\theequation}{S\arabic{equation}}
\renewcommand{\thesection}{S\arabic{section}}
\tableofcontents
\hfill \\
\twocolumngrid

\section{Hopf invariant defined via inverse images}
\label{sec.gapped}

The most general form of a 2-band complex Hamiltonian in three-dimension is given as
\begin{gather}
    H(\bk) = \mathbf{a}(\bk)\cdot\bm{\sigma} \n
    = a_1(\bk)\sigma_{x} + a_2(\bk)\sigma_{y} + a_3(\bk)\sigma_{z}.
\label{eq.2band}
\end{gather}
To topologically classify the 2-band Hamiltonian, we consider the energetically flattened Hamiltonian where the energy of the occupied state is $-1$ and that of the unoccupied state is $+1$ without changing eigenstates.
Then the flattened Hamiltonian is fully described by the normalized vector $\hat{\mathbf{a}}\in\mathbb{S}^{2}$, which is called by the Bloch vector, and the sphere where the Bloch vector lives is called by the Bloch sphere.
With the trivial Chern class condition on all three subtori in the Brillouin zone, the Bloch vector is represented by a map from $\mathbb{S}^3$ to $\mathbb{S}^2$, which allows us to topologically classify the Hamiltonian by the third homotopy group $\pi_{3}(\mathbb{S}^{2})=\mathbb{Z}$.
This integer-valued invariant is the Hopf invariant $\chi$ which is given by
\begin{gather}
\chi=-\frac{1}{4{\pi}^2}\int_{\mathrm{BZ}} d^{3}k ~\mathbf{F}\cdot \mathbf{A}\in\mathbb{Z},
\end{gather}
where $\mathbf{A}$ and $\mathbf{F}$ are the abelian Berry connection and the Berry curvature of the occupied state, respectively~\cite{wen2008hopf,deng2013hopf}.

\begin{figure}[t]
\includegraphics[width=\linewidth]{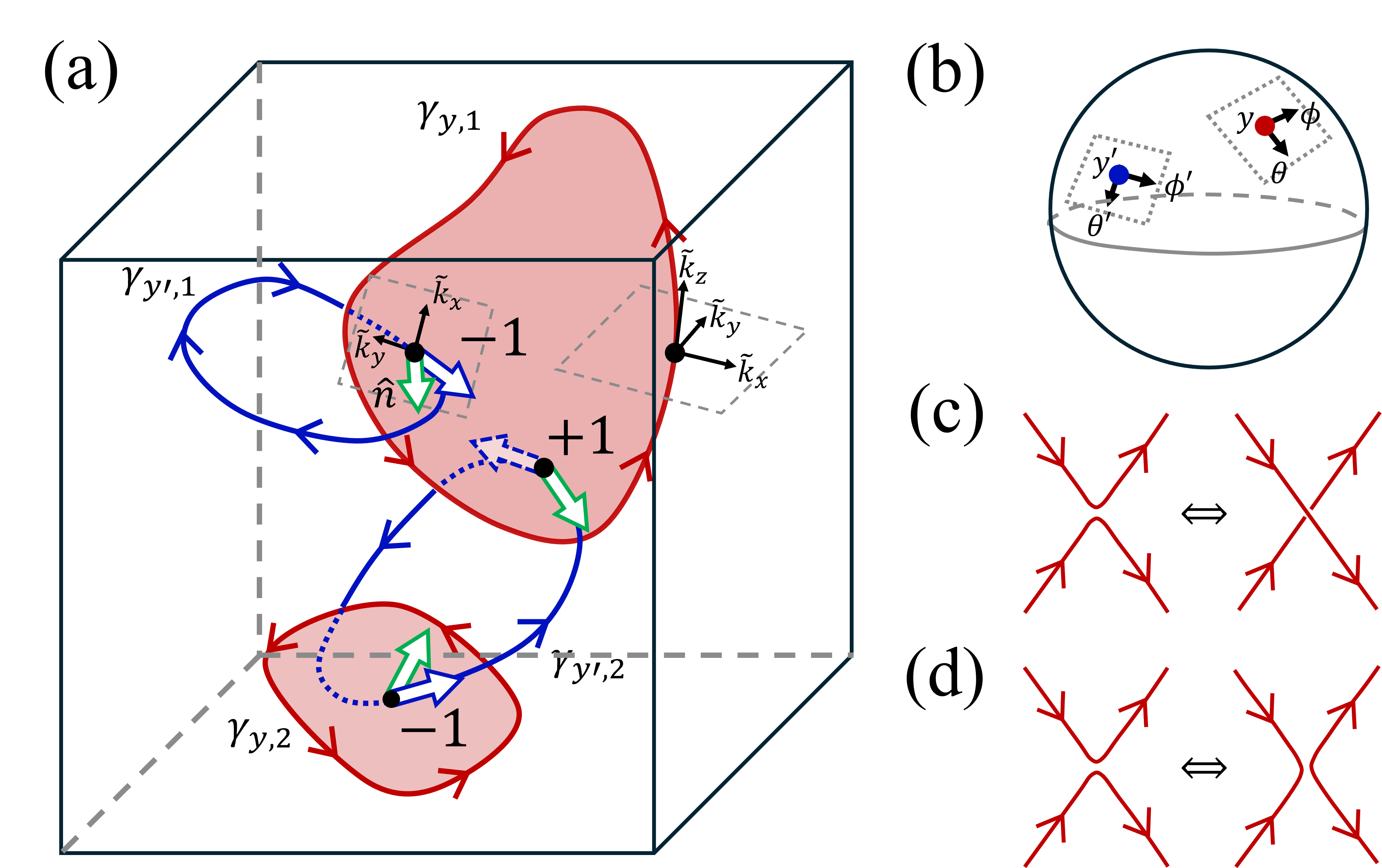}
\caption{Depiction of inverse images of the Hopf insulator. (a) A schematic figure which shows two inverse images $\gamma_y=\gamma_{y,1}\cup\gamma_{y,2}$ (red) and $\gamma_{y'}=\gamma_{y',1}\cup\gamma_{y',2}$ (blue) of two different points $y$ and $y'$ on the Bloch sphere. 
Utilizing a local coordinate $(\tilde{k}_x,\tilde{k}_y,\tilde{k}_z)$, it can be shown that the Berry curvature vector at an arbitrary point on $\gamma_{y}$ is tangential with the loop.
The surface $\Sigma_{y}$ (colored by red) for $\gamma_{y}$ is assigned by respecting the orientation of $\gamma_{y}$, which is defined by the opposite direction with the Berry curvature.
For counting the linking number between $\gamma_{y}$ and $\gamma_{y'}$, one can investigate the sign of an inner product between the normal vector of $\Sigma_{y}$ (green arrow) and the tangential vector of $\gamma_{y'}$ (blue arrow) at the points where $\gamma_{y'}$ pierces $\Sigma_{y}$.
Another local coordinate at the piercing point is illustrated to support the proof connecting the Whitehead's formula and this method of counting the linking number.
(b) Depiction of two points $y$ and $y'$ and the coordinates $(\theta,\phi)$, $(\theta',\phi')$ on the Bloch sphere. (c), (d) Illustrations of some typical merging behavior between the inverse images.
}
\label{Fig2_0420}
\end{figure}

In particular, Hopf invariant can be formulated by introducing the inverse image $\gamma_y=h^{-1}[{y}]$ of the flattened Hamiltonian $h:\mathbb{S}^3 \rightarrow \mathbb{S}^2$ for an arbitrary fixed point $y\in\mathbb{S}^2$ on the Bloch sphere. Generically, $\gamma_y$ in the 3D Brillouin torus is a collection of one-dimensional closed loops $\gamma_{y,i}$, since they cannot have an open end due to the continuity of the Hamiltonian. In order to give the orientation to each loop, we first prove that the Berry curvature vector at a single point $X\in\gamma_{y,i}$ is always parallel with $\gamma_{y,i}$ at the point.
The Berry curvature for the 2-band Hamiltonian defined in Eq. (\ref{eq.2band}) can be represented as
\begin{gather}
    \mathrm{F}_{k}=\sum_{i,j}\varepsilon_{ijk}{\sin{\theta}\over2}{\partial(\theta,\phi)\over\partial(k_{i},k_{j})},
\label{eq.Berry2band}
\end{gather}
where $(\theta,\phi)$ is the spherical coordinates on the Bloch sphere, and $\varepsilon_{ijk}$ is the Levi-Civita symbol. Now considering a point $X\in\gamma_{i}$ and the basis of a locally rotated Cartesian coordinate $(\tilde{k}_{x},\tilde{k}_{y},\tilde{k}_{z})$ with $\mathbf{\tilde{k}_{z}}$ direction defined parallel to $\gamma_{i}$ through $X$ (see Fig. \ref{Fig2_0420} (a) and (b)), it is clear that the parallel component  $\mathrm{F}_{\tilde{k}_{z}}$ of the Berry curvature only survives because $\theta$ and $\phi$ are constant along $\mathbf{\tilde{k}_{z}}$.
Thus, the orientation of $\gamma_{i}$ can be well-defined as the opposite of the Berry curvature direction. We also note that this orientation cannot be flipped along $\gamma_{i}$ as long as it is in the shape of a simple loop: Berry curvature should be 0 at some point on $\gamma_{i}$ for the orientation flip, but then one can find a particular direction that $(\theta,\phi)$ coordinate is preserved in a plane that is normal to the direction of $\gamma_{i}$ at the point. This leads to the need for another inverse image loop crossing the point.

If we assign a set of oriented surfaces, $\{\Sigma_{i}\}$, which satisfies the condition $\partial\left(\cup_{i} \Sigma_{i}\right)=\gamma$, Hopf invariant $\chi$ is given by Whitehead's integral formulation \cite{whitehead1947expression,
nelson2021multicellularity}
\begin{gather}
    \chi=\sum_{i}{{1 \over 2\pi}\int_{\Sigma_{i}}  \bf{F}\cdot d{\boldsymbol\Sigma}}.
\label{eq.Whitehead}
\end{gather}
This can be interpreted as the sum of well-defined Chern numbers $C_{i}$ on each $\Sigma_{i}$ since one can topologically contract the boundary $\gamma_{i}$ to a point on the Bloch sphere and regard $\Sigma_{i}$ as a closed surface.

Furthermore, Whitehead's formulation directly leads to the linking number examination of the Hopf invariant. The image of $\Sigma_{i}$ wraps the Bloch sphere as the same number as the Chern number $C_{i}$. Accordingly, the number which $\gamma'$ pierces $\Sigma_{i}$ must be the same as $C_{i}$, when considering the inverse image $\gamma'=\cup_{i}\gamma'_{i}$ of an arbitrary different point $y'\neq y$ in the Bloch sphere. Note that one should assign a proper sign respecting the orientation of local diffeomorphism when counting the number that $\gamma'$ pierces $\Sigma_{i}$: Consider again a local coordinate $(\tilde{k}_{x},\tilde{k}_{y},\tilde{k}_{z})$  around a piercing point which is defined such that one basis vector $\mathbf{\tilde{k}_{z}}$ is the same as the normal vector $\mathbf{n}$ of $\Sigma_{i}$. The local diffeomorphism is orientation preserving (reversing) if the Jacobian ${\partial(\theta,\phi)}/{\partial(\tilde{k}_{x},\tilde{k}_{y})}$ is positive (negative), which has a same sign as the Berry curvature component normal to the surface $\mathrm{F}_{\tilde{k}_{z}}=\mathbf{F}\cdot{\mathbf{n}}$ at a point Eq. (\ref{eq.Berry2band}). Given that $\gamma'$ is always antiparallel to the Berry curvature, counting the occurrence of piercing while considering the orientation of each $\gamma_{i}$ reduces to the issue of counting the linking number between $\gamma$ and $\gamma'$.

Lastly in this section, we discuss the possibility of touching between two loops of inverse images. Let us make several remarks: (a) Touching between different inverse images $\gamma$ and $\gamma'$ of the different points on the Bloch sphere $y$ and $y'$ is forbidden in the insulating phase. If we assume there exists a touching point $\mathbf{k}_{0}$ of $\gamma$ and $\gamma'$, the image $h(\mathbf{k}_{0})$ at the point should map to both $y$ and $y'$, which means the Bloch sphere is no longer well-defined. This can only happen when the gap closing is happening at $h(\mathbf{k}_{0})$. (b) Touching and merging between the loops $\gamma_{i}$ and $\gamma_{j}$ which are the different parts of the inverse image of a fixed point $y$ is allowed, and occurs if and only if the Berry curvature $\mathbf{F}=0$ at the touching point. For the only if part, we already have shown it when discussing the statement that the orientation cannot be flipped along the single loop. For the if part, it is because the direction of the Berry curvature vector at the touching point should not be specified in order to hold two generically different orientations of inverse images at the same point. (c) Merging between $\gamma_{i}$ and $\gamma_{j}$ takes place in the manner of preserving global orientations of inverse images. Consider that the Hamiltonian $H(\mathbf{k},\phi)$ is continuously changing over the time parameter $\phi$, and inverse image loops are also dynamically varying the shape. Comparing the infinitesimal time before and after the merging, the global shape and the orientation of the inverse images need to be preserved except at the very local vicinity around the touching point. Some illustrations of the allowed merging behavior are demonstrated in Fig. \ref{Fig2_0420} (c) and (d). (d) This dynamical movement including the merging of inverse images cannot change the total linking number, which is simply the consequence of the remark (c). This is consistent and also works as a prerequisite for the fact that the Hopf invariant should be well-defined in the insulating phase. Practical examples that show the merging between the inverse images would be further demonstrated in Sec. \ref{sec.processes}.

\section{Weyl point trajectories and the Hopf insulator phase transition}
\label{sec.processes}
\begin{figure}[t]
\includegraphics[width=\linewidth]{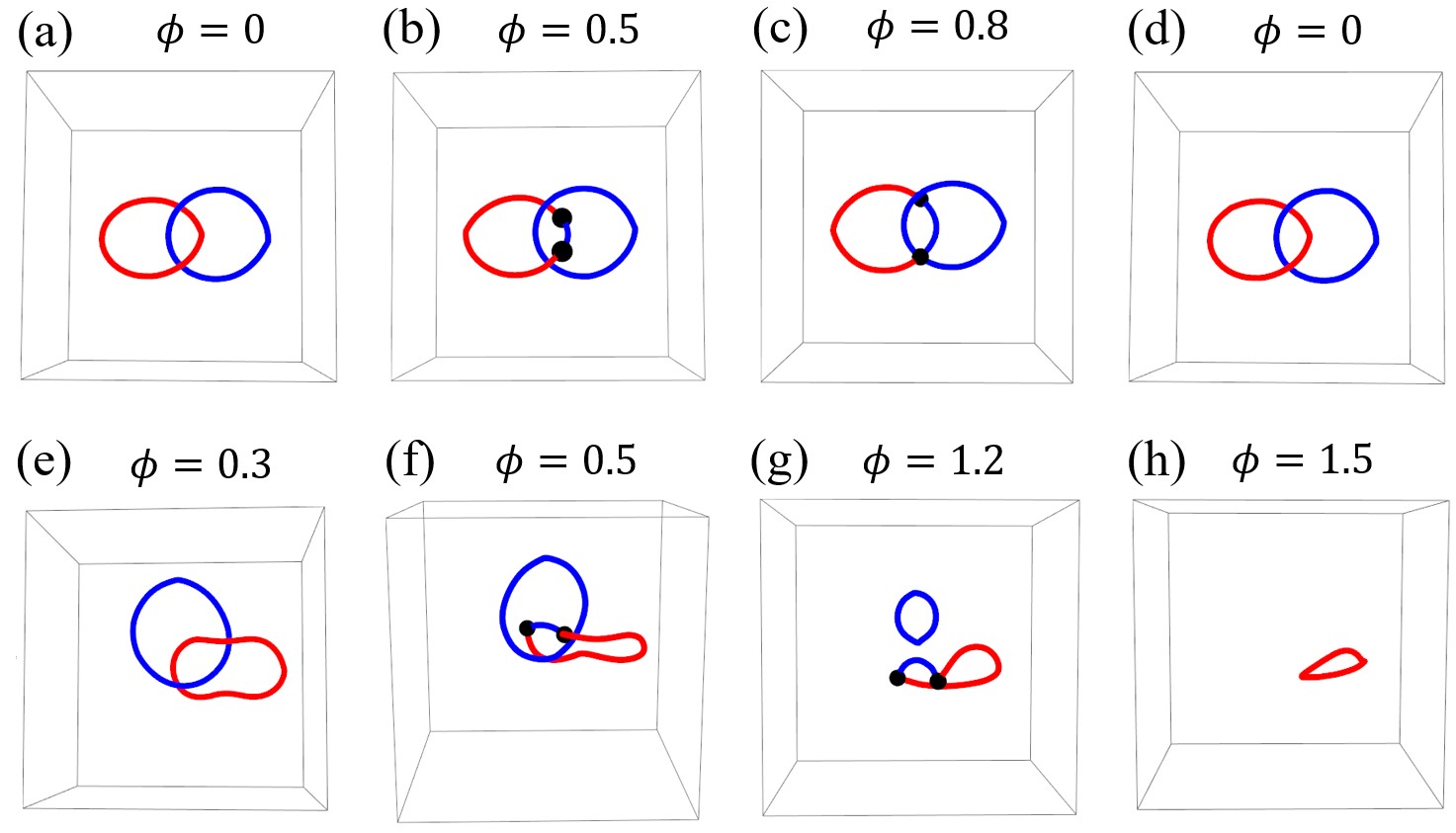}
\caption{
Two phase transitions of the Hopf insulator with the same trajectory of the Weyl point (Black dot) and one inverse image of a given point on the Bloch sphere (Red line), but different variation of the Hopf invariant.
(a)-(d) The Weyl point and the inverse images of the points on the Bloch sphere for the model in Eq.~(\ref{WPT1}) while $\phi$ changes from $0$ to $0.8$ to $0$.
The red loop (resp. blue loop) is the inverse image of $(-1,0,0)$ (resp. $(1,0,0)$) on the Bloch sphere.
(e)-(h) The Weyl point and the inverse images of the points on the Bloch sphere for the model in Eq.~(\ref{WPT2}) while $\phi$ changes from $0.3$ to $1.5$.
The red loop (resp. blue loop) is the inverse image of $(0,\frac{\sqrt{3}}{2},-\frac{1}{2})$ (resp. $(1,0,0)$) on the Bloch sphere.}
\label{fig:WPT}
\end{figure}

In this section, we consider two Hopf insulator models undergoing the gapless phase to show that the trajectory of Weyl points is insufficient to classify the topological phase transition of the Hopf insulator.
The first model is given by:
\begin{gather}
z=\begin{pmatrix}
\sin k_{x} +i\sin k_{y} 
\\ \sin{k_{z}}+i\left(\sum_{i=x,y,z} \cos{k_{i}} +\frac{5}{2}\right)
\end{pmatrix}
\\
\mathbf{a}=z^{\dagger}\boldsymbol{\sigma}z+\phi\hat{x},~~H=\mathbf{a}\cdot\boldsymbol{\sigma},
\label{WPT1}
\end{gather}
where $\boldsymbol{\sigma}=\sigma_{x}\hat{x}+\sigma_{y}\hat{y}+\sigma_{z}\hat{z}$ is the vector of the Pauli matrices. 
Here the parameter $\phi$ changes from $0$ to $0.8$ to $0$, and, through this process, we track not only the gap closing point, illustrated the black dots in Fig. \ref{fig:WPT} (a)-(d), 
but also the inverse image of the point $(-1,0,0)$ on the Bloch sphere, illustrated the red line in Fig. \ref{fig:WPT} (a)-(d).
In this process, a pair of Weyl points are pair-created and then pair-annihilated themselves.
The red loop in $\phi=0$ becomes an open line after pair-creation of the Weyl points and then returns to the closed loop after the pair-annihilation.
During this process, the change of the Hopf invariant is zero because the initial and the final Hamiltonians are the same.

The second model is given by:
\begin{gather}
z=\begin{pmatrix}
\sin k_{x} +i\sin k_{y} 
\\ \sin{k_{z}}+i\left(\sum_{i=x,y,z} \cos{k_{i}} +2+\phi \right)
\end{pmatrix}
\\
\mathbf{a}=z^{\dagger}\boldsymbol{\sigma}z+\phi\hat{x},~~H=\mathbf{a}\cdot\boldsymbol{\sigma},
\label{WPT2}
\end{gather}
Here, we increases the parameter $\phi$ from $0.3$ to $1.5$, and, through this process, we track not only the gap closing point, the black dots in Fig. \ref{fig:WPT} (e)-(h),
but also the inverse image of the point $(0,\frac{\sqrt{3}}{2},-\frac{1}{2})$ on the Bloch sphere, the red line in Fig. \ref{fig:WPT} (e)-(h).
Similar to the first model, while increasing $\phi$, the pair-created Weyl points are pair-annihilated themselves and the red loop becomes an open line and then closed loop.
However, unlike the first model, the Hopf invariant of the second model changes from one to zero.
This difference becomes clear if we follow the inverse image of the another point on the Bloch sphere.
For the both model, we see the inverse image of $(1,0,0)$ on the Bloch sphere, which is the blue line in Fig. \ref{fig:WPT}.
In the case of the first model in Eq.~(\ref{WPT1}), one can find that the linking between the blue loop and the red loop do not change through the gapless phase.
In the case of the second model in Eq.~(\ref{WPT2}), however, the linking number changes through the gapless phase by passing the blue loop through the blue open line attached to the Weyl points.
Therefore, we cannot specify the phase transition of the Hopf insulator by following the Weyl point trajectory and the inverse image of one point on the Bloch sphere.

\section{Predictions for other Altland-Zirnbauer classes}
%
In this section, we give a more formal analysis of the unconventional topological phase transition (TPT) of three-dimensional (3D) topological band systems. First, employing the mathematical theory of ``universal 2-covering spaces," we explain why the TPT of Hopf insulator cannot be fully described by the Weyl point trajectory, in stark contrast with its low-dimensional cousin, namely, Chern insulator. Next, we exemplify the mathematical principle by the `real Hopf insulator' in symmetry class AI, whose phase transition is expected to differ from the 2D Euler insulator in a similar manner to how Hopf and Chern insulators do. Finally, we expand the analysis to the whole tenfold symmetry classes, proposing all candidates of 3D topological insulators whose restriction on low-dimensional topology demands that its TPT should be described in view of the eigenstates $\psi(\bk)$ rather than the projector $P(\bk) = \sum_i \ket{\psi(\bk)}\bra{\psi(\bk)}$, as is the case for Hopf insulator.

\subsection{Formal view on Hopf insulator phase transition}

Since the two-band topological insulators in class A has the classifying space
\begin{gather}
    \mathfrak{X}_{\mathrm{A}} = \frac{\mathrm{U}(2)}{\mathrm{U}(1)\times \mathrm{U}(1)} \simeq \mathbb{S}^{2},
    \label{clsfofHopf}
\end{gather}
any such insulator can be identified with a continuous map
\begin{gather}
    f: \mathbb{T}^{d}_{\text{BZ}} \to \mathbb{S}^{2}
\end{gather}
where $\mathbb{T}^{d}_{\text{BZ}}$ stands for the $d$-dimensional Brillouin Zone (BZ) torus.
In spatial dimensions $d=3$, the Hopf insulator is defined under the condition that the Chern numbers are trivial ($C=0$) on all three faces of the BZ torus. At the same time, a Hamiltonian describing a Hopf insulator corresponds to a mapping $\mathbb{T}^{3}_{\text{BZ}} \to \mathbb{S}^3 \rightarrow \mathbb{S}^2$ that ``factors through" the 3-sphere $\mathbb{S}^{3}$, while a general complex 2-band Hamiltonian corresponds to a general mapping $\mathbb{T}^3_{\text{BZ}} \rightarrow \mathbb{S}^2$ that does not necessarily factor through $\mathbb{S}^{3}$. 
It turns out that the intermediate space $\bb{S}^3$ is the set of (normalized) occupied eigenstates of a 2 by 2 Hamiltonian, so the factor map $\mathbb{T}^3_{\text{BZ}} \rightarrow \mathbb{S}^3$ continuously and globally assigns occupied eigenstate $\psi(\bk)$ to each $\bk$. Since this assignment is impossible unless the Chern numbers vanish, it is consistent with the previously mentioned condition that all Chern numbers defined on 2d surfaces of $\mathbb{T}^3_{\text{BZ}}$ must be trivial. We will see in Sec.~\ref{supp.n-cover} that the vanishing of Chern numbers is also sufficient for the existence of factoring $\mathbb{T}^{3}_{\text{BZ}} \to \mathbb{S}^3 \rightarrow \mathbb{S}^2$.

The situation is succinctly summarized by the notion of \textit{lifting}:
\begin{gather}
\label{cd.hopf}
\begin{tikzcd}[ampersand replacement=\&] 
    \& \mathbb{S}^{3} \dar["p"] \\
    \mathbb{T}^{3}_{\text{BZ}} \ar[ur,"\tilde{f}"] \rar["f"] \& \mathbb{S}^{2}
\end{tikzcd}
\end{gather} 
While general 2-band 3D insulator in class A is represented by a map $f: \mathbb{T}^{3}_{\text{BZ}} \to \mathbb{S}^{2}$, Hopf insulator enjoys the property that the map $f$ may be lifted to another map $\tilde{f}$ that points upstairs in the diagram \eq{cd.hopf}. That $\tilde{f}$ is a lift of $f$ is equivalent to saying that $p\circ\tilde{f} = f$.

The vertical map $p$, called Hopf fibration, maps a circle $\mathbb{S}^{1}_{x}$ onto a single point $x\in\mathbb{S}^{2}$. Therefore, finding a lift $\tilde{f}$ of a general map $f$ is equivalent to choosing a point $\widetilde{f}(\bk) \in \mathbb{S}^{1}_{f(\bk)}$ coherently for all momenta $\bk\in\mathbb{T}^{3}_{\text{BZ}}$. The obstruction for finding such a lift is precisely the nonzero Chern numbers on $(k_x,k_y),\, (k_y,k_z),\,(k_z,k_x)$ reduced BZ tori $\mathbb{T}^{2}_{\mathrm{rBZ}}$.

The upshot is that a generic 2-band Hamiltonian may be parametrized as
\begin{subequations}
    \begin{gather}
    H(\bk) = f_x(\bk) \sigma_x + f_y(\bk) \sigma_y + f_z(\bk) \sigma_z, 
    \\ 
    f = (f_x, f_y, f_z): \mathbb{T}^{3}_{\text{BZ}} \to \mathbb{R}^{3}, 
    \\
    f_x^2 + f_y^2 + f_z^2 = 1,
\end{gather}
\end{subequations}
Hopf insulator is naturally parametrized as in
\begin{subequations}
    \begin{gather}
    H_{\text{Hopf}}(\bk) = -[\tilde{f}^{\dagger}(\bk) \bm{\sigma} \tilde{f}(\bk)] \cdot\bm{\sigma}, 
\label{eq.hopf}
    \\ 
    \tilde{f}: \mathbb{T}^{3}_{\text{BZ}} \to \mathbb{C}^{2}, 
    \\ 
    \|\tilde{f}\|^2 = 1.
\end{gather}
\end{subequations}
In fact, the vector $\tilde{f}(\bk)$ is the occupied eigenstate of $H_{\text{Hopf}}(\bk)$. Hence, the lifted map $\tilde{f}$ contains the full spectral data of the Hamiltonian.
The Hopf fibration $p$ is precisely the map
\begin{gather}
    p: \mathbb{S}^{3} \to \mathbb{S}^{2} \n 
    \tilde{f} \mapsto -\tilde{f}^{\dagger} \bm{\sigma} \tilde{f} = (f_x, f_y, f_z),
\end{gather}
which discards the information of phase angle,
\begin{gather}
    p(\tilde{f}e^{i\theta}) = p(\tilde{f}),
\end{gather}
and hence can be interpreted as the projector function
\begin{gather}
    p: \mathbb{S}^{3} \to \mathbb{S}^{2} \n 
    \ket{\psi} \mapsto \ket{\psi}\bra{\psi} = P(\bk).
\end{gather}

Regarding the phase transition, a stark difference between $H$ and $H_{\text{Hopf}}$ is the following: One needs to solve four independent equations
\begin{gather}
    \Re[\tilde{f}(\bk)] = 0, \quad
    \Im[\tilde{f}(\bk)] = 0, 
\label{eq.4eqs}
\end{gather}
to find a gapless point of Hopf insulator $H_{\text{Hopf}}$, while only three equations
\begin{gather}
    f_x(\bk) = 0, \quad f_y(\bk) = 0, \quad f_z(\bk) = 0
\label{eq.3eqs}
\end{gather}
are needed for the generic Hamiltonian $H$ (note that \eq{eq.4eqs} harbors four independent equations since $\tilde{f}(\bk)$ is a two-dimensional vector). Hence, in the four dimensional parameter space $\mathbb{T}^{3}_{\text{BZ}}\times \mathbb{R}_{\lambda}$ where $\lambda\in\mathbb{R}_{\lambda}$ is a tunable parameter in the Hamiltonian, the gapless points form a zero $(4-4=0)$ dimensional manifold, i.e. discrete points. In other words, the Hamiltonian $H_{\text{Hopf}}(\bk, \lambda)$ undergoes a direct transition, where the gapless state appears only at discrete values of $\lambda$, not supporting a gapless phase (where gapless state appears for some interval $a\leq \lambda \leq b$). In contrast, gapless points of the general Hamiltonian $H(\bk,\lambda)$ form a one $(4-3=1)$ dimensional manifold, which contains a continuous interval of $\lambda$ and supports a stable gapless phase.

In fact, one can perturb the Hamiltonian, 
\begin{gather}
    H_{\text{Hopf}}(\bk,\lambda) \to (H_{\text{Hopf}} + \delta H)(\bk,\lambda),
\end{gather}
so that the perturbed Hamiltonian has a stable gapless phase. However, our discussions on the unperturbed Hamiltonian $H_{\text{Hopf}}$ has a significant implication for the phase transition of the perturbed ones. Namely, the trajectory of gapless Weyl points may not be sufficient to describe the change of topological charge; we already have an instance of model, $H_{\text{Hopf}}(\bk,\lambda)$, that describes phase transition of Hopf insulator but contains no such thing as Weyl point trajectory other than the trivial `trajectory' (single point). This allows us to conclude that unlike general topological insulators in class A, describing the phase transition of Hopf insulator involves more information (namely, information of individual states $\psi(\bk)$) than Weyl point trajectory (which is fully determined by the projector $P(\bk) = \ket{\psi(\bk)} \bra{\psi(\bk)}$). Such unconventional nature of TPT is a consequence of the restriction that the Hamiltonian is trivial on 2D restricted Brillouin zones.

%
\subsection{Universal $n$-covering spaces and (Real-)Hopf insulator
\label{supp.n-cover}}
%
To generalize the discussion in the previous subsection to other symmetry classes, let us introduce the notion of universal $n$-covering spaces. It will soon be clear that the case $n=2$ is the most relevant for our analysis of unconventional TPT.

A continuous map $p: \widetilde{X}^{(n)} \to X$ is called a \textit{universal $n$-covering} of $X$ if the following conditions are met:

\begin{subequations}
\label{eqs.ncover}
\begin{align}
    &\text{For all } i\leq n, \quad \pi_i[\widetilde{X}^{(n)}] = 0. 
\label{eq.ncover_1}
    \\ 
    &\text{For all } i>n, \quad p_{*}: \pi_i[\widetilde{X}^{(n)}] \xrightarrow[]{\sim} \pi_i[X].
\label{eq.ncover_2}
\end{align}
\end{subequations}

In \eq{eq.ncover_2}, $p_{*}$ denotes the group homomorphism induced by $p$ on $i$-th homotopy groups. The condition requires that it is a group isomorphism for every $i>n$. When \eq{eqs.ncover} is satisfied, it is also said that $\widetilde{X}^{(n)}$ is a universal $n$-covering space, or $n$-connected cover~\cite{MR1867354}, of $X$. 

It is a standard topic in algebraic topology that the universal $1$-covering, often abbreviated as universal cover, translates a lifting problem (as in \eq{cd.hopf}) into homotopy theory and vice versa. More precisely, when $p: \widetilde{X}^{(1)} \to X$ is a universal $1$-covering,
\begin{gather}
\label{cd.1cover}
\begin{tikzcd}[ampersand replacement=\&] 
    \& \widetilde{X}^{(1)} \dar["p"] \\
    Y \ar[ur,"\tilde{f}"] \rar["f"] \& X
\end{tikzcd}
\end{gather} 
a map $f:Y\to X$ admits a lift $\tilde{f}: Y\to \widetilde{X}^{(1)}$ if and only if the induced homomorphism on the first homotopy group $\pi_1$ is zero:
\begin{gather}
    0 = f_{*} : \pi_1[Y] \to \pi_1[X].
\end{gather}
For example, a model of Su-Schrieffer-Heeger chain is equivalent to a classifying map $f_{\text{SSH}}: \mathbb{T}^{1}_{\text{BZ}} \to X$ where $X$ is a relevant classifying space. The chain is topologically trivial if and only if the Berry phase is $0$ mod $2\pi$. Since $f$ induces the map
\begin{gather}
    f_{\text{SSH}*}: \pi_1[\mathbb{T}^{1}] \simeq \mathbb{Z} \to \mathbb{Z}_{2} \simeq \pi_1[X]
\end{gather}
where $A(k)$ is the Berry connection.
Depending on realization of the group $\bb{Z}_{2}$, we have
\begin{subequations}
\begin{alignat}{2}
    & f_{\text{SSH}*}(n) =  \exp{in\int_{0}^{2\pi} dk A(k)} \quad 
    &&(\bb{Z}_{2} = \{\pm1\}), 
    \\ 
    & f_{\text{SSH}*}(n) = \frac{n}{\pi} \int_{0}^{2\pi} dk A(k) \text{ mod } 2 \quad 
    &&(\bb{Z}_{2} = \{0,1\}).
\label{supp.ssh_map}
\end{alignat}
\end{subequations}
Due to symmetry condition, the Berry phase is quantized:
\begin{gather}
    \int_{0}^{2\pi} dk A(k) = 
    \begin{cases}
        0 \mod 2\pi & (\text{trivial}) \\
        \pi \mod 2\pi & (\text{topological})
    \end{cases}.
\end{gather}
As a result, the chain is topologically trivial with zero Berry phase if and only if $f_{\text{SSH}*} = 0$, in the notation of \eq{supp.ssh_map}. Equivalently, the chain is topologically trivial if and only if $f_{\text{SSH}}$ lifts to some
\begin{gather}
    \tilde{f}_{\text{SSH}}: \mathbb{T}^{1}_{\text{BZ}} \to \widetilde{X}^{(1)},
\end{gather}
that is, $f_{\text{SSH}}$ factors through $\widetilde{X}^{(1)}$.

Likewise, a universal $n$-covering bridges the lifting problem and homotopy theory, but involves higher homotopy groups:
\begin{gather}
\label{cd.ncover}
\begin{tikzcd}[ampersand replacement=\&] 
    \& \widetilde{X}^{(n)} \dar["p"] \\
    Y \ar[ur,"\tilde{f}"] \rar["f"] \& X
\end{tikzcd}
\end{gather} 
In \eq{cd.ncover}, the map $f$ admits a lift if and only if it induces trivial homomorphisms on all homotopy groups $\pi_i$ for $i\leq n$:
\begin{gather}
    0 = f_{*} : \pi_i[Y] \to \pi_i[X] \quad \text{for all } i\leq n.
\label{eq.ncover_obs}
\end{gather}
We observe that the universal 2-covering is relevant to the Hopf insulator: Since 3D Hopf insulator restricts to trivial 2D insulators on any of the planes $(k_x,k_y),\, (k_y,k_z),\, (k_z,k_x)$, the map $f_{\text{Hopf}}$ representing a Hopf insulator induces the trivial homomorphisms
\begin{gather}
    0 = f_{\text{Hopf}*}: \pi_i[\mathbb{T}^{3}] \to \pi_i[\mathfrak{X}_{\mathrm{A}}] \quad \text{for all } i\leq 2.
\end{gather}
Therefore, a map $f:\mathbb{T}^{3} \to \mathfrak{X}_{\mathrm{A}}$ represents a Hopf insulator if and only if it lifts to $\widetilde{\mathfrak{X}}_{\mathrm{A}}^{(2)}$. It turns out that 
\begin{gather}
    \mathfrak{X}_{\mathrm{A}} \simeq \mathbb{S}^{2}, \quad 
    \widetilde{\mathfrak{X}}_{\mathrm{A}}^{(2)} \simeq \mathbb{S}^{3},
\label{eq.2cover_hopf}
\end{gather}
reproducing the claims in the previous subsection.

Before proceeding, let us give another example of universal 2-covering that appears in the analysis of topological insulator. Band systems with spinless $\mathcal{PT}$ symmetry, where $\mathcal{P},\,(\mathcal{T})$ is the space inversion (time reversal), are represented by the maps
\begin{gather}
    f: \mathbb{T}^{d}_{\text{BZ}} \to \mathfrak{X}_{\mathrm{AI}} \simeq \frac{\mathrm{O}(m+n)}{\mathrm{O}(m)\times \mathrm{O}(n)}.
\end{gather}
For two occupied $(m=2)$ and two unoccupied $(n=2)$ bands, the classifying space admits familiar universal 1- and 2-coverings:
\begin{subequations}
\begin{align}
    \mathfrak{X}_{\mathrm{AI}} &\simeq \frac{\mathrm{O}(4)}{\mathrm{O}(2)\times \mathrm{O}(2)}, 
    \\    
    \widetilde{\mathfrak{X}}_{\mathrm{AI}}^{(1)}
     &\simeq \mathbb{S}^{2}\times \mathbb{S}^{2}, 
    \\
    \widetilde{\mathfrak{X}}_{\mathrm{AI}}^{(2)}
     &\simeq \mathbb{S}^{3}\times \mathbb{S}^{3}.
\end{align}
\end{subequations}
Therefore, using the Hopf map construction $f_{\text{Hopf}}: \mathbb{T}^{3}_{\text{BZ}}\to \mathbb{S}^{3} \to \mathbb{S}^{2}$, one can construct
\begin{gather}
    f_{\text{RHopf}}: \mathbb{T}^{3}_{\text{BZ}} \to \mathbb{S}^{3}\times\mathbb{S}^{3} \to \mathbb{S}^{2}\times\mathbb{S}^{2}
\end{gather}
and study the band systems represented by $f_{\text{RHopf}}$. Such band insulators are called Real Hopf insulator~\cite{lim2023realhopf}.

%
\subsection{Tenfold generalization of Hopf insulator phase transition}
%

\begin{table*}[t]
    \centering
    \begin{tabular}{c | c| c| c| c| c| c| c| c| c| c}
    \hline\hline
    Class & A & AI & AII & AIII & BDI & CII & D & C & DIII & CI
    \\ \hline
    $G$ 
    & $\mathrm{U}(m+n)$ 
    & $\mathrm{O}(m+n)$ 
    & $\mathrm{Sp}(m+n)$ 
    & $\mathrm{U}(n)\cdot \mathrm{U}(n)$
    & $\mathrm{O}(n)\cdot \mathrm{O}(n)$
    & $\mathrm{Sp}(n)\cdot \mathrm{Sp}(n)$
    & $\mathrm{O}(2n)$
    & $\mathrm{Sp}(n)$
    & $\mathrm{U}(2n)$
    & $\mathrm{U}(n)$
    \\
    $H$ 
    & $\mathrm{U}(m)\cdot \mathrm{U}(n)$
    & $\mathrm{O}(m)\cdot \mathrm{O}(n)$
    & $\mathrm{Sp}(m)\cdot \mathrm{Sp}(n)$
    & $\mathrm{U}(n)$ 
    & $\mathrm{O}(n)$ 
    & $\mathrm{Sp}(n)$ 
    & $\mathrm{U}(n)$
    & $\mathrm{U}(n)$
    & $\mathrm{Sp}(n)$
    & $\mathrm{O}(n)$
    \\
    \hline\hline
    \end{tabular}
    \caption{Lie groups that appear as a factor in classifying spaces $\mathfrak{X}=G/H$ of tenfold symmetry classes due to Altland-Zirnbauer. $m$ is the number of unoccupied bands and $n$ is the number of occupied bands. In classes that have chiral (AIII, BDI, CII) or particle-hole (D, C, DIII, CI) symmetries, $m=n$ and hence the groups depend on $n$ alone. Once the classifying space $\mathfrak{X}=G/H$ is fixed, the $d$-dimensional Hamiltonians are represented by continuous maps $f: \mathbb{T}^{d}_{\text{BZ}} \to \mathfrak{X}$.}
\label{tab:AZ_groups}
\end{table*}

We propose that the universal 2-covering is responsible for the unique characteristics of Hopf insulator phase transition, which is not describable solely in terms of the Weyl point trajectory (or any other information encoded in the projector $P=\ket{\psi}\bra{\psi}$). In fact, what was needed for the full description of the phase transition process is the knowledge of individual eigenstates $\psi(\bk)$, whose phase angle is completely forgotten by the projector $P(\bk)$. At the same time, we found that the universal 2-covering space in \eq{eq.2cover_hopf} relevant to the Hopf insulator was actually the space of Bloch eigenstates, while the classifying space itself is the space of projectors, carrying strictly less information than the former. In this regard, it was inevitable that the covering space $\widetilde{\mathfrak{X}}_{\mathrm{A}}^{(2)}$ had larger dimensions than $\mathfrak{X}_{\mathrm{A}}$, resulting in more equations to solve to find the gap closing conditions; see Eqs.~\eqref{eq.4eqs}-\eqref{eq.3eqs}.

In the rest of this section, we generalize the analysis of Hopf insulator to all Altland-Zirnbauer (AZ) symmetry classes. To this end, we observe that every classifying space of AZ symmetry class takes the form of
\begin{gather}
    \mathfrak{X} = G/H,
\end{gather}
where $G$ is the configuration space of Bloch eigenstates, and $H$ is the gauge group. An action of $h\in H$ on $G$ mixes the occupied (unoccupied) eigenstates, yielding the same projector
\begin{gather}
    P = \sum_{i: \text{occupied}} \ket{\psi_i} \bra{\psi_i}.
\end{gather}
Therefore, $\mathfrak{X}=G/H$ serves as the space of projectors.

Now, a naive idea to generalize unconventional TPT of Hopf insulator to tenfold classes is to look for all band systems for which the following is true:
\begin{gather}
    G \to G/H = \mathfrak{X}
    \quad \text{is a universal 2-cover.}
\label{supp.naive_hypothesis}
\end{gather}
However, the postulate in \eq{supp.naive_hypothesis} needs to be properly modified. Before turning to such details, let us discuss the ideas behind this postulate.

The choice of 2-covering space, among the general $n$-covering spaces, is the following. For $n=1$, it is a general fact that the universal cover $\widetilde{X}^{(1)}$ of a space $X$ has the same dimension as $X$. It is only for $n\geq 2$ that the universal $n$-cover $\widetilde{X}^{(n)}$ is expected to have larger dimensions than $X$, 
which signals that $\widetilde{X}^{(n)}$ contains more information that $X$. For two-band systems such as Hopf insulator, this increase of dimensions would imply stronger gap closing conditions, which is observed from Eqs.~\eqref{eq.4eqs}-\eqref{eq.3eqs}. In contrast, the number of gap closing equations for general $N$-band systems is not directly related to the dimension of $\widetilde{X}^{(n)}$.

In any case, $\widetilde{X}^{(n)}$ having larger dimensions than $X$ will result in a qualitatively different descriptions of phase transition for general $(f)$ and lifted $(\tilde{f})$ Hamiltonians. Specifically, phase transition of the latter will require the full information of Bloch eigenstates, $\psi\in G$, while the knowledge of projectors $P\in G/H$ would suffice for the former.

Furthermore, due to the properties of $n$-covering spaces discussed in Eqs.~\eqref{cd.ncover}-\eqref{eq.ncover_obs}, the non-trivial low dimensional homotopy classes
\begin{align}
    0\neq \operatorname{im} f_{*}\Big|_{\pi_i[\mathbb{T}^{d}]} &\equiv (\text{image of } f_{*} \text{ on } i\text{-th homotopy}) 
    \n 
    &\subseteq \pi_i[\mathfrak{X}] \quad (i\leq n)
\end{align}
are exactly the obstruction to the existence of lifting $\tilde{f}$. Physically, this means that $\tilde{f}$ represents the special class of Hamiltonians that are trivial in $n$-dimensional sub-tori of $d$-dimensional BZ. This naturally demands $d\gneqq n$. However, unless we are concerned with Floquet systems, $d=3$ is the largest dimension that can be found in realistic materials. Hence, we restrict our analysis to the case $n=2$. Accordingly, all the entries in our classification will be filled with 3D materials.

To see if $G \to \mathfrak{X}$ qualifies as a universal 2-cover, we look at the exact sequence~\cite{MR1867354} 

\begin{gather}
\label{cd.LES_master}
\begin{tikzcd}[ampersand replacement=\&] 
    \cdots \to \pi_{3}[H] \rar["i_{*}"] \& 
    \pi_{3}[G] \rar["p_{*}"] \dar[ phantom, ""{coordinate, name=Z3}] \& \pi_{3}[\mathfrak{X}] \ar[dll,"\delta_{*}" near start, rounded corners, to path={ -- ([xshift=3ex]\tikztostart.east)
    |- (Z3) [near end]\tikztonodes
    -| ([xshift=-3ex]\tikztotarget.west)
    -- (\tikztotarget)}] \\ 
    \pi_2[H] \rar["i_{*}"] \& 
    \pi_2[G] \rar["p_{*}"] \dar[ phantom, ""{coordinate, name=Z2}] 
    \& \pi_2[\mathfrak{X}] \ar[dll,"\delta_{*}" near start, rounded corners, to path={ -- ([xshift=3ex]\tikztostart.east)
    |- (Z2) [near end]\tikztonodes
    -| ([xshift=-3ex]\tikztotarget.west)
    -- (\tikztotarget)}]  \\ 
    \pi_1[H] \rar["i_{*}"] \& 
    \pi_1[G] \rar["p_{*}"]
    \& \pi_1[\mathfrak{X}] \to \cdots.
\end{tikzcd}
\end{gather} 

In \eq{cd.LES_master}, the arrows represent the group homomorphisms induced on homotopy groups by the canonical inclusion $(i)$ and projection $(p)$ maps:
\begin{gather}
    i: H\to G, \quad p: G\to \mathfrak{X}.
\end{gather}
$\delta_{*}$ is called boundary map, whose detailed definition is not important for our discussion. The important property of exact sequence is that, at each entry between a pair of arrows, we have
\begin{gather}
    \operatorname{im}(g) = \ker(h)\quad \text{in} \quad A \xrightarrow[]{g} B \xrightarrow[]{h} C.
\label{eq.exactness}
\end{gather}
In particular, $A,\,B,\,C$ may be some of the homotopy groups in \eq{cd.LES_master}, and $g,\,h$ may be $i_{*},\,p_{*},\,\delta_{*}$.

For $p: G\to \mathfrak{X}$ to be a universal 2-covering, two conditions in \eq{eqs.ncover} should be met. In view of the exact sequence, \eq{cd.LES_master}, these conditions can be restated as

\begin{subequations}
    \label{eqs.2cover_int}
\begin{align}
    &\pi_1[G] = 0 \quad\text{and}\quad \pi_2[G] = 0. 
\label{eq.2cover_int1} \\ 
    &\pi_i[H]=0 \quad\text{for all} \quad i\geq 3.
\label{eq.2cover_int2} 
\end{align}
\end{subequations}
To see that \eq{eq.2cover_int2} is equivalent to \eq{eq.ncover_2} for 2-cover $(n=2)$, note that the following patterns are repeated in \eq{cd.LES_master}:
\begin{subequations}
    \label{eqs.patterns}
\begin{gather}
    \pi_d[H] \xrightarrow{i_{*}} 
    \pi_d[G] \xrightarrow{p_{*}} 
    \pi_d[\mathfrak{X}] \xrightarrow{\delta_{*}} 
    \pi_{d-1}[H]
\label{eq.pattern_1}
    \\
    \pi_{d+1}[\mathfrak{X}] \xrightarrow{\delta_{*}} 
    \pi_{d}[H] \xrightarrow{i_{*}} 
    \pi_d[G]
\label{eq.pattern_2}
\end{gather}
\end{subequations}
Applying the exactness condition in \eq{eq.exactness} to \eq{eq.pattern_1}, it follows that $p_{*}$ is an isomorphism if and only if the adjacent homomorphisms vanish, $i_{*}=\delta_{*}=0$. On the other hand, exactness of \eq{eq.pattern_2} implies that $\pi_d[H]=0$ if and only if the adjacent homomorphisms vanish, $i_{*}=\delta_{*}=0$. Now, suppose \eq{eq.ncover_2} holds. Then by exactness of Eqs.~\eqref{eqs.patterns}, $p_{*}$ being isomorphism for all $d\geq 3$ implies $\pi_{d}[H]=0\, (\forall d\geq 3).$ Conversely, suppose $\pi_{d}[H]=0\, (\forall d\geq 3).$ Since every Lie group has trivial second homotopy group~\cite{Bott1956, bryan1994texture}. We have $\pi_{2}[H]=0$ as well. Again, exactness of Eqs.~\eqref{eqs.patterns} implies that $p_{*}$ is an isomorphism for all $d\geq 3$.

However, for physical considerations, we modify the conditions in Eqs.~\eqref{eqs.2cover_int} into
\begin{subequations}
\label{eqs.2cover_mod}
\begin{align}
    & \pi_2[G] = 0 \quad\text{while}\quad \pi_2[\mathfrak{X}]\neq 0. 
\label{eq.2cover_mod1} \\ 
    & \pi_i[H]=0 \quad\text{for all} \quad i\geq 3.
\label{eq.2cover_mod2} 
\end{align}
\end{subequations}
Note that the condition $\pi_1[G]=0$ is dropped and replaced by $\pi_2[\mathfrak{X}]\neq 0$.

The reasoning behind the modification is the following. If $\pi_1[G]\neq 0$, one can substitute its universal 1-cover $\widetilde{G}^{(1)}$ for $G$ itself. Generally, the 1-cover $\widetilde{X}^{(1)}$ of a space $X$ has equal dimensions as $X$, in fact, they are locally homeomorphic. As a consequence, both $\widetilde{G}^{(1)}$ and $G$ encodes (locally) the same information about the eigenstates, and thus we propose that they can be used interchangeably. Then, the property of 1-covering space guarantees that
\begin{subequations}
\begin{align}
    \pi_1[\widetilde{G}^{(1)}] &= 0,
    \\
    \pi_{*}: \pi_2[\widetilde{G}^{(1)}] \xrightarrow{\;\sim\;} \pi_2[G] &= 0,
\end{align}
\end{subequations}
where $\pi: \widetilde{G}^{(1)} \to G$ is the 1-covering map. Furthermore, the composite homomorphisms
\begin{gather}
    (p\circ \pi)_{*}: \pi_i[\widetilde{G}^{(1)}] \xrightarrow{\;\sim\;} \pi_i [G] \xrightarrow{\;\sim\;} \pi_i [\mathfrak{X}] 
\end{gather}
are isomorphisms for $i>2$. Hence, $\widetilde{G}^{(1)}$ not only contains the information about Bloch eigenstates, but also qualifies as a universal 2-cover of the classifying space $\mathfrak{X}$. This perfectly meets our purpose, which ultimately justifies discarding the condition $\pi_1[G]=0$.

Turning to the other modification in \eq{eq.2cover_mod1}, we require $\pi_2[\mathfrak{X}]\neq 0$ in order to have non-trivial obstruction for lifting; otherwise, any classifying map $f$ will satisfy the vanishing condition in a trivial way,
\begin{gather}
    0 = f_{*}: \pi_2[\bb{T}^{3}] \to \pi_2[\mathfrak{X}] = 0,
\end{gather}
rendering our analysis rather uninteresting. It is only when $\pi_2[\mathfrak{X}]\neq 0$ that there is a special class of Hamiltonians inducing trivial homomorphism 
\begin{gather}
    0 = f_{*}: \pi_2[\mathbb{T}^{3}] \to \pi_2[\mathfrak{X}] \neq 0,
\end{gather}
whose phase transition is distinguished from the general Hamiltonian $(f_{*}|_{\pi_2[\mathbb{T}^{3}]}\neq 0)$, thus being ``unconventional".

Using the exactness property, \eq{eq.exactness}, the non-triviality condition $\pi_2[\mathfrak{X}]\neq 0$ is equivalent to
\begin{gather}
    0\neq \ker i_{*}: \pi_1[H] \to \pi_1[G].
\label{eq.non-triv}
\end{gather}
To see this, note that $\pi_2[G]=0$ since $G$ is a Lie group. Then by exactness at $\pi_2[\mathfrak{X}]$ we have
\begin{gather}
    \pi_2[\mathfrak{X}] 
    \simeq \operatorname{im} \delta_{*} \big|_{\pi_2[\mathfrak{X}]} 
    \simeq \ker i_{*} \big|_{\pi_1[H]},
\label{eq.purelykernel}
\end{gather}
from which \eq{eq.non-triv} follows.

Therefore, we are to tabulate all 3D insulators whose classifying space $\mathfrak{X} = G/H$ satisfies

\begin{subequations}
    \label{eqs.2cover_final}
\begin{align}
    \ker i_{*}\Big|_{\pi_1[H]} & \neq 0 \quad ( 
    i: H \hookrightarrow G),
\label{eq.2cover_final1} \\ 
    \pi_i[H] & = 0 \quad (\forall i\geq 3).
\label{eq.2cover_final2}
\end{align}
\end{subequations}

The table of groups $G,\,H$ for each AZ symmetry class is listed in Table~\ref{tab:AZ_groups}. Among them, \eq{eq.2cover_final1} is satisfied only for the five classes
\begin{gather}
    \mathrm{A,\, AI,\, D,\, C,\, CI}.
\end{gather}
The other condition, \eq{eq.2cover_final2}, restricts the possible number of bands:
\begin{subequations}
\begin{align}
    \mathrm{A} &: m = n = 1, \\
    \mathrm{AI} &: m, n \leq 2, \\
    \mathrm{D} &: n = 1, \\
    \mathrm{C} &: n = 1, \\
    \mathrm{CI} &: n \leq 2.
\label{eq.fivefold_pre}
\end{align}
\end{subequations}
This is because $\pi_3[\mathrm{U}(n)] = \mathbb{Z}$ for all $n\geq 2$ and $\pi_3[\mathrm{O}(n)] = \mathbb{Z}$ for $n\geq 3$.

In particular, since there is a strong constraint on the number of both occupied and unoccupied bands, we conclude that the change of phase transition due to the increased number of gap closing conditions is a characteristic of \textit{delicate} topological insulators, whose topology is protected only under a constraint on the number of total bands.

Finally, we can discard the trivial 3D insulators $(\pi_3[G/H]=0)$ in the list \eq{eq.fivefold_pre}. After that, we have the complete list of nontrivial 3D insulators,
\begin{gather}
    \textbf{Final Result.} \n
\begin{subequations}
    \label{supp.fourfold_res_2}
\begin{align}
    \mathrm{A} &: (m,n) = (1,1), \\
    \mathrm{AI} &: (m,n) = (1, 2),\, (2, 1),\, (2,2), \\
    \mathrm{C} &: (m,n) = (1,1), \\
    \mathrm{CI} &: (m,n) = (2,2),
    \label{supp.fourfold_res_3}
\end{align}
\end{subequations}
\end{gather}
whose phase transition should be described in terms of eigenstates $\ket{\psi}$ instead of the projector $P=\ket{\psi}\bra{\psi}$. Note that the unique case in class A is the Hopf insulator, and the case $(m,n)=(2,2)$ in class AI corresponds to the real Hopf insulator. We defer to later works the detailed investigation of unconventional TPT of the systems listed in \eq{supp.fourfold_res_2}.

\section{Three-dimensional topological invariants of the classes in Eqs.(\ref{supp.fourfold_res_2} - \ref{supp.fourfold_res_3})}

In this section, we investigate the three-dimensional topological invariants of the classes in Eqs.(\ref{supp.fourfold_res_2} - \ref{supp.fourfold_res_3}) and show that they can be commonly reduced back to a form of the Hopf invariant.
Thus, the form of their topological invariants illustrates that the unconventional TPT, in the sense of this manuscript, is deeply related to the Hopf map.

First, the class A with $(m,n)=(1,1)$ is the Hopf insulator and it has three-dimensional topological invariant as the Hopf invariant~\cite{wen2008hopf}.
The origin of the Hopf invariant is based on the isomorphism between its classifying space and $\mathbb{S}^2$, explained in Eq. (\ref{clsfofHopf}).

Next, let us see the class AI with $(m,n)=(2,1)$.
From Table. \ref{tab:AZ_groups}, one can find that its classifying space $\mathfrak{X}_{\mathrm{AI},(2,1)}$ is given by
\begin{gather}
\mathfrak{X}_{\mathrm{AI},(2,1)}=\frac{\mathrm{O}(3)}{\mathrm{O}(2)\times\mathrm{O}(1)}\simeq \mathbb{S}^{2}/\mathbb{Z}_{2},
\end{gather}
where we use that $\mathrm{O(3)/O(2)}\simeq\mathbb{S}^{2}$, and $\mathrm{O}(1)\simeq\mathbb{Z}_{2}$.
Since the second and third homotopy groups of $\mathbb{Z}_{2}$ is trivial, the third homotopy group of $\mathfrak{X}_{\mathrm{AI},(2,1)}$ is the same as that of $\mathbb{S}^{2}$ and the three-dimensional topological invariant can be calculated using the Hopf invariant.

To figure out how to calculate the three-dimensional topological invariant of $\mathfrak{X}_{\mathrm{AI},(2,1)}$, we need to focus on its occupied eigenstate $|u_{\mathrm{AI},(2,1)}^{\mathrm{occ}}\rangle$.
Since there are total three bands in this class, $|u_{\mathrm{AI},(2,1)}^{\mathrm{occ}}\rangle$ is a real-valued three-component vector, which means that $|u_{\mathrm{AI},(2,1)}^{\mathrm{occ}}\rangle\in \mathbb{S}^{2}$.
Hence, we can calculate the topological invariant by regarding each component of $|u_{\mathrm{AI},(2,1)}^{\mathrm{occ}}\rangle$ as $a_{i} (i=1,2,3)$ in Eq. (\ref{eq.2band}) and calculating the Hopf invariant of the Hamiltonian $H=\mathbf{a}\cdot\bm{\sigma}$.
In this case, the Bloch sphere is given by the occupied eigenstate space. 
It should be noted that the three-dimensional topological invariant of the class AI with $(m,n)=(1,2)$ is determined by applying the same formula as the class AI with $(m,n)=(2,1)$ but using the unoccupied state, not the occupied state.
In the case of the class AI with $(m,n)=(2,2)$, its classifying space $\mathfrak{X}_{\mathrm{AI},(2,2)}$ is given by \cite{lim2023realhopf}
\begin{gather}
\mathfrak{X}_{\mathrm{AI},(2,2)}=\frac{\mathrm{O(4)}}{\mathrm{O(2)}\times\mathrm{O(2)}}\simeq\frac{\mathbb{S}^{2}\times\mathbb{S}^{2}}{\mathbb{Z}_{2}},
\end{gather}
which means that the three-dimensional topological invariants are given by two Hopf invariants.
This class is known as the real Hopf insulator and explained in detail in the paper \cite{lim2023realhopf}.

In the case of the class C with $(m,n)=(1,1)$, it has the anti-unitary symmetry $\mathfrak{P}$ which is anti-commute with the Hamiltonian and satisfies $\mathfrak{P}^{2}=-1$ \cite{bzdusek2017robust}.
In the case of the two-band Hamiltonian, this is automatically achieved by setting $\mathfrak{P}=i\sigma_{y}\mathcal{K}$ which anti-commutes with all the Pauli matrices, where $\mathcal{K}$ is the complex-conjugation operator.
Therefore, the classifying space of the class C with $(m,n)=(1,1)$ is exactly the same as that of the class A with $(m,n)=(1,1)$.

Finally, let us see the class CI with $(m,n)=(2,2)$ whose classifying space $\mathfrak{X}_{\mathrm{CI},(2,2)}$ is given by
\begin{gather}
\mathfrak{X}_{\mathrm{CI},(2,2)}=\mathrm{U}(2)/\mathrm{O}(2).
\end{gather}
Since the second and third homotopy groups of the orthogonal group $\mathrm{O}(2)$ are trivial, the third homotopy group of $\mathfrak{X}_{\mathrm{CI},(2,2)}$ is the same as that of $\mathrm{U(2)}$.
According to the paper \cite{lapierre2021nband}, the three-dimensional topological invariant of $\mathrm{U(2)}$ is given by the Hopf invariant.
Therefore, the three-dimensional topological invariant of $\mathfrak{X}_{\mathrm{CI},(2,2)}$ has a form of the Hopf invariant.

To figure out how to calculate the three-dimensional topological invariant of $\mathfrak{X}_{\mathrm{CI},(2,2)}$, let us determine the symmetry representations.
The class CI has the space-time inversion symmetry $\mathfrak{T}$ and the chiral symmetry $\mathcal{C}$ which commute to each other \cite{bzdusek2017robust}.
In the case of the class CI with $(m,n)=(2,2)$, we choose their representations by
\begin{gather}
\mathfrak{T}=\mathcal{K},~\mathcal{C}=\begin{pmatrix}
1&0&0&0\\
0&1&0&0\\
0&0&-1&0\\
0&0&0&-1
\end{pmatrix}.
\end{gather}
Then the two occupied states $|u_{i}^{occ}\rangle$ and two unoccupied states $|u_{i}^{unocc}\rangle$, where $i=1,2$, are given by
\begin{gather}
|u_{i}^{occ}\rangle=\begin{pmatrix}
|u_{i}^{\uparrow}\rangle\\
|u_{i}^{\downarrow}\rangle
\end{pmatrix},~
|u_{i}^{unocc}\rangle=\begin{pmatrix}
|u_{i}^{\uparrow}\rangle\\
-|u_{i}^{\downarrow}\rangle
\end{pmatrix},
\end{gather}
where $i=1,2$ and $|u_{i}^{\uparrow}\rangle$, and $|u_{i}^{\downarrow}\rangle$ are real-valued two-component vectors.
Considering the orthonormal condition of $|u_{i}^{occ}\rangle$ and $|u_{i}^{unocc}\rangle$, one can find that the two complex-valued two-component vectors $|u_{1}^{c}\rangle$, $|u_{2}^{c}\rangle$, which are given by
\begin{gather}
|u_{1}^{c}\rangle=|u_{1}^{\uparrow}\rangle+i|u_{1}^{\downarrow}\rangle,~
|u_{2}^{c}\rangle=|u_{2}^{\uparrow}\rangle+i|u_{2}^{\downarrow}\rangle,
\end{gather}
form a basis of $\mathbb{C}^{2}$ space.
Therefore, the Hopf invariant calculated using $|u_{1}^{c}\rangle$ is the three-dimensional topological invariant of the class.